\documentclass{aastex631}
\usepackage{amsmath}

\received{April 1, 2025}
\revised{October 31, 2025}
\revised{November 31, 2025}
\accepted{December 6, 2025}

\submitjournal{AJ}
\graphicspath{{./}}

\begin{document}

\title{Metrics for Optimizing Searches for Orbital Precession and Tidal Decay via Transit- and Occultation-Timing}

\author[0000-0002-9495-9700]{Brian Jackson}
\affiliation{Department of Physics, Boise State University, 1910 University Drive, Boise ID 83725-1570 USA}
\affiliation{Carl Sagan Center, SETI Institute, Mountain View, CA, United States}
\author[0000-0002-9131-5969]{Elisabeth R.~Adams}
\affiliation{Planetary Science Institute, 1700 E. Ft. Lowell, Suite 106, Tucson, AZ 85719, USA}
\author[0000-0002-3548-1655]{Rachel M.~Huchmala}
\affiliation{Department of Physics, Boise State University, 1910 University Drive, Boise ID 83725-1570 USA}
\affiliation{Schmid College of Science and Technology, Chapman University, Orange, CA, USA}
\author[0000-0002-9495-9700]{Malia Barker}
\affiliation{Department of Physics, Boise State University, 1910 University Drive, Boise ID 83725-1570 USA}
\affiliation{Department of Computer Science, Boise State University, 1910 University Drive, Boise ID 83725-1570 USA}
\author[0009-0004-5054-664X]{Marvin Rothmeier}
\affiliation{Institute of Space Systems, University of Stuttgart, Pfaffenwaldring 29, 70569 Stuttgart, Germany}

\author[0000-0003-3716-3455]{Jeffrey P.~Morgenthaler}
\affiliation{Planetary Science Institute, 1700 E. Ft. Lowell, Suite 106, Tucson, AZ 85719, USA}
\author[0000-0002-9468-7477]{Amanda A.~Sickafoose}
\affiliation{Planetary Science Institute, 1700 E. Ft. Lowell, Suite 106, Tucson, AZ 85719, USA}

\begin{abstract}
Short-period exoplanets may exhibit orbital precession driven by several different processes, including tidal interactions with their host stars and secular interactions with additional planets. This motion manifests as periodic shifts in the timing between transits which may be detectable via high-precision and long-baseline transit- and occultation-timing measurements. Detecting precession and attributing it to a particular process may constrain the tidal responses of planets and point to the presence of otherwise undetected perturbers. However, over relatively short timescales, orbital decay driven by the same tidal interactions can induce transit-timing signals similar to the precession signal, and distinguishing between the two processes requires robust assessment of the model statistics. In this context, occultation observations can help distinguish the two signals, but determining the precision and scheduling of observations sufficient to meaningfully contribute can be complicated. In this study, we expand on earlier work focused on searches for tidal decay to map out simple metrics that facilitate detection of precession and how to distinguish it from tidal decay. We discuss properties for a short-period exoplanet system that can maximize the likelihood for detecting such signals and prospects for contributions from citizen-science observations.
\end{abstract}


\section{Introduction} \label{sec:Introduction}
Increasingly long observational baselines and a burgeoning planetary assemblage are revealing subtler and yet more important aspects of exoplanetary systems. For no planets is this more true than transiting exoplanets. Transits for some exoplanets have been observed for more than a decade now. Campaigns to monitor radial velocity excursions indicative of short-period (periods $\lesssim$ few days) planets began in earnest in the mid-1990s as new technology allowed for higher precision measurements \citep{1994SPIE.2198..362V, 1996A&AS..119..373B}, and a complete transit of the planet HD 209458 b, the first of any exoplanet, was observed using sub-meter-class telescopes and was reported in 2000 \citep{2000ApJ...529L..45C, 2000ApJ...529L..41H}. Transit observations of HD 209458 b continue even today, with one of the most recent, a ground-based observation, successfully detecting the subtle spectral signatures of water and ammonia \citep{2022A&A...657A..23E}.

The simplicity, symmetry, and strength of transit signals mean the moment at the center of the transit, called the mid-transit time, can be estimated to impressive precision. For example, WASP-12 b, another member of the hot Jupiter class, was discovered in 2009 \citep{2009ApJ...693.1920H}, and since then, hundreds of transits have been observed, both from the ground and from space. \citet{2021AJ....161...72T} used lightcurves collected in space by NASA's TESS mission and determined the planet's mid-transit times to precisions better than 50-s. 

Such transit-timing precisions, coupled with decades-long baselines, can facilitate detection of even very small departures from a completely periodic, Keplerian orbit. Here, again, WASP-12 b serves as a key example. Suggestions that WASP-12 b might exhibit non-Keplerian motion go back at least to 2011 when \citet{2011ApJ...727..125C} analyzed eclipses of the planet observed by the Spitzer Space Telescope and proposed the planet's orbit was precessing. However, as transit upon transit was observed, the precision of the planet's orbital ephemeris improved dramatically so that, by 2018, its orbital period had been determined as $1.09142172\pm0.00000015\,{\rm days}$, a precision better than 13 milliseconds \citep{2018AcA....68..371M}. By that point, the departure from a Keplerian orbit had become clear, but whether that departure arose from tidal decay or orbital precession remained unclear. 

Both processes had been proposed to impact the orbits of short-period exoplanets going back to their discovery. Indeed, the first exoplanet around a Sun-like star, 51 Peg b \citep{1995Natur.378..355M}, garnered concerns about orbital stability against tidal effects \citep{1996ApJ...470.1187R}. Orbiting their host stars within several stellar and planetary radii, short-period exoplanets experience and induce significant distortions from tidal gravity. For a non-zero orbital eccentricity and/or rotation states not synchronized to the orbital motion, tides raised on the star or planet can drive orbital circularization and decay. In terms of transit-timing, this decay manifests as a small but steady decrease in the orbital period over many years as the planet spirals into the host star. 

\begin{figure}
    \centering
    \includegraphics[width=\linewidth]{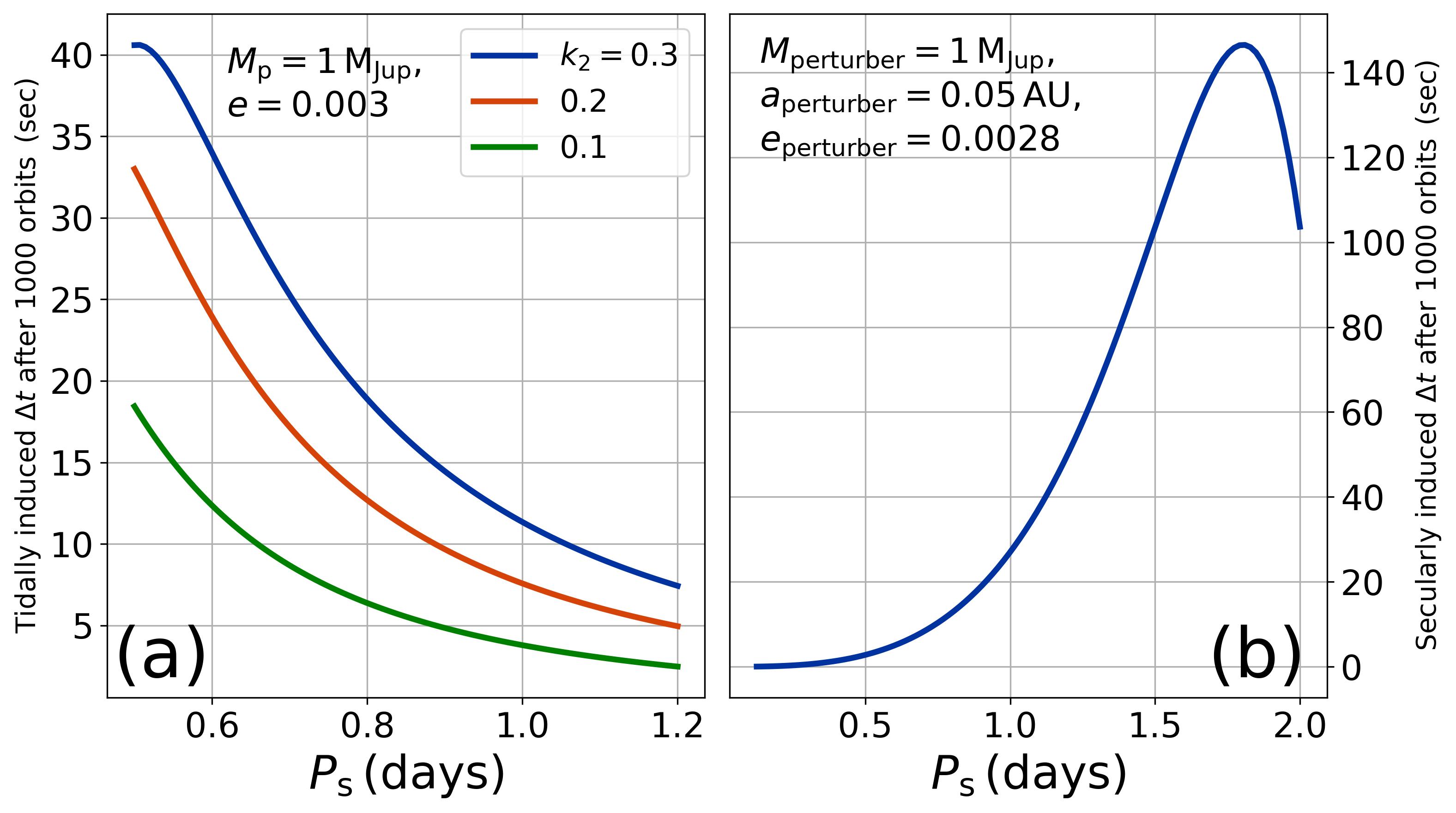}
    \caption{Examples of precessionally driven transit-timing departures, $\Delta t$, from a strictly periodic Keplerian orbit. (a) Precession driven by tidally induced distortion of a short-period gas giant with a sidereal period $P_{\rm s}$, an assumed mass $M_{\rm p} = 1$ Jupiter mass (${\rm M_{Jup}}$), an orbital eccentricity $e = 0.003$, and a range of Love numbers $k_2$. The Love number depends on the planet's internal structure and tidal response -- see \citet{2009ApJ...698.1778R} for additional details. (b) Precession for the smaller, inner planet driven by secular interactions in a two-planet, WASP-47-like system consisting of a $9$-Earth mass planet with a sidereal period $P_{\rm s}$ interior to a co-planar, $1\,{\rm M_{Jup}}$ planet with a semi-major axis $a = 0.05\,{\rm AU}$ and orbital eccentricity $e = 0.0028$, all around a star with a mass equal to the Sun's. Here, we have neglected tidal damping, which would act over much longer timescales than shown. The size of $\Delta t$ is insensitive to the interior planet's orbital eccentricity, as long as it is not identically zero.}
    \label{fig:Example_Precession_Rates}
\end{figure}

For non-zero orbital eccentricities, the same tidal distortions can also induce orbital precession, and one form of such precession involves revolution of the orbital major axis, called apsidal precession. As the major axis revolves, the time from one transit to the next periodically lengthens or shortens, depending on the orbit's orientation to the observer's line-of-sight to the system. \citet{2009ApJ...698.1778R} considered this form of precession and pointed out that the rate of precession was governed, in part, by the amplitude of the planet's tidal distortion which itself depends on the planet's internal structure as parameterized by the Love number $k_2$. Therefore, the rate of tidally induced precession could constrain interior structures. Using the model outlined in \citet{2009ApJ...698.1778R}, Figure \ref{fig:Example_Precession_Rates}(a) shows examples of how much precession would be expected after 1,000 orbits for a short-period gas giant planet. Even for the small eccentricity considered, a Jupiter-like planet might exhibit tens of seconds of precession over just a few years, rendering it easily detectable with long enough baselines and sufficiently precise transit-timing data.

The presence of additional perturbing bodies in the system can also drive precession, also leading to transit-timing variations. Of course, transit-timing variations arising from mean-motion resonances (MMRs) have been used for years to constrain orbital architectures and planetary masses \citep[e.g.,][]{2016ApJS..225....9H}. However, even non-resonant multi-body gravitational interactions can drive precession that manifests in transit-timing variations; MMRs just drive especially large variations. Such variations, if attributable to multi-body interactions, could be leveraged to infer the presence of additional otherwise undetected planets. Figure \ref{fig:Example_Precession_Rates}(b) shows the scale of transit-timing variations in a system with short-period, Earth-like planet in an orbit with precession driven by a nearby and co-planar Jupiter-like planet, meant to mimic a WASP-47-like system \citep{2023A&A...673A..42N}. For this calculation, we solved the Laplace-Lagrange first-order secular equations \citep[cf.,][]{1999ssd..book.....M} and then calculated the change in transit-timing using Equation \ref{eqn:transit_precession_ephemeris} as shown below. We can see, again, this precession could result in measurable transit-timing variations. 

Investigating such secularly-driven transit-timing variations might corroborate ideas regarding the orbital architectures of systems tightly packed inner planets STIPs. \citet{2023AJ....166...36H} recently showed that STIPs with massive outer planets seem to exhibit more complex gaps between the planets' orbits than STIPs without massive outer planets, where the complexity was quantified using an information theory approach \citep{2020AJ....159..281G}. \citet{2025ApJ...979..202L} suggested that this increased complexity could be explained if orbital inclinations among the STIPs are secularly driven to large enough values that not all the planets can be seen transiting from Earth. Searches for long-term secularly-driven transit-timing variations among such systems could test this hypothesis of present, yet-unseen planets.

Returning to WASP-12 b, a concerted effort found no signs of additional planets in the system \citep{2017AJ....153...78C}, and after more than ten years of transit observations, \citet{2020ApJ...888L...5Y} definitively concluded the planet is truly spiraling into its host star and not precessing. However, other systems show at least tentative evidence for precession and not tidal decay. For example, \citet{2022AJ....163..208B} suggested the possibility that the hot Jupiter HAT-P-30 b/WASP-51 b exhibits precession on the basis of a few dozen transits, many observed by citizen scientist astronomers using amateur-grade telescopes. More recently, \citet{2025AA...694A.233B} suggested that WASP-43 b might be experiencing both orbital decay and precession.

Distinguishing between tidal decay, which manifests as a monotonic decrease in the times between transits, and orbital precession, which manifests as a sinusoidal variation in those times, is particularly challenging since over a short enough timescale the precession signal can resemble the decay signal, at least when only transits are considered. A tool that has come into common use to decide which model provides the best fit to timing data is the Bayesian Information Criterion or ${\rm BIC}$ \citep{1978AnSta...6..461S}. Under certain reasonable simplifying assumptions, ${\rm BIC}$ can be written as
\begin{equation}
    {\rm BIC} = \chi^2 + k \ln N,
\end{equation}
where $\chi^2$ is the sum of scaled residuals \citep{1997iea..book.....T}, $k$ is the number of parameters used to fit the desired model, and $N$ is the number of data points to which the model is fit. ${\rm BIC}$ provides a balance between finding a model that fits a given dataset ($\chi^2$ small) and choosing a parsimonious model that includes a minimal number of model parameters. For instance, the model for transit-timing variations driven by precession (Equation \ref{eqn:transit_precession_ephemeris}) involves five fit parameters ($k = 5$), while the tidal decay model involves only three ($k = 3$). Therefore, the precession model will often return a smaller $\chi^2$, but the addition of the penalty term in ${\rm BIC}$ for introducing more model parameters means tidal decay can sometimes be preferred. When comparing two low-dimension ($k$ small) models, the model giving the smaller ${\rm BIC}$ is said to be preferred. Often, this comparison is cast in terms of $\Delta{\rm BIC} = {\rm BIC}_{\rm A} - {\rm BIC}_{\rm B}$, the difference between model A and model B. Cast this way, $\Delta{\rm BIC} > 0$ implies a preference for model B, while $\Delta{\rm BIC} < 0$ implies a preference for model A. 

However, ${\rm BIC}$ involves some important limitations and caveats \citep[cf.,][]{24ce203a-855a-3aa9-952f-976d23b28943}. For instance, $\ln\left({\rm BIC} \right)$ is meant to approximate the Bayes factor (i.e., the relative evidence in favor of one model against another) but has an error $\sim O(1)$. Also, ${\rm BIC}$ is governed by the same statistical variability that governs the behavior of $\chi^2$. In other words, statistical fluctuations in a dataset may drive variability in ${\rm BIC}$. As we show below, this variability means small but non-zero values of $\Delta{\rm BIC}$ do not necessarily imply a preference for one model over another, and the variability scales with the number of data points. We discuss in this study how to leverage expectations from ${\rm BIC}$ statistics to make quantitative predictions for how future timing observations would affect the confirmation of tidal decay or precession in an exoplanetary system.

Even with a voluminous transit-timing dataset in hand, distinguishing one kind of non-Keplerian orbital motion from another may be challenging. Even though WASP-12 b had 102 transit-timing data points spanning more than nine years, \citet{2017AJ....154....4P} indicated tidal decay was only marginally preferred over precession. A key development between 2017 and 2020 was the collection of  occultation-timing data: it was the addition of five occultation-timing data as reported in \citet{2020ApJ...888L...5Y} that helped settle the case for tidal decay in the WASP-12 system. As discussed below, tidal decay and precession drive occultation-timing in opposite directions. The decline in orbital period due to tidal decay shortens the times between subsequent transits and between subsequent occultations. However, precession will shorten the time between one set of times (either transits or occultations) while lengthening the time between the other set of times. Thus, combining timing data for transits and occultations provides a uniquely potent means for distinguishing tidal decay and precession. Moreover, occultation data are also key for distinguishing other processes that impact timing, including barycentric acceleration, though, in such cases, radial velocity data are also important \citep{2020ApJ...888L...5Y}.

Thus, a successful program of transit observations for exoplanets with the goal of detecting and accurately attributing timing variations requires coordination of numerous observational resources over long timescales. Both observations already completed and proposed future observational campaigns, as well as the astrophysical properties of a given exoplanetary system, should all be folded into developing a plan for detecting non-Keplerian orbital motion. In this study, we develop several simple metrics for optimizing a search for such orbital evolution using timing data. This study expands on a previous study, \citet{2023AJ....166..142J}, which focused exclusively on tidal decay. For the present study, we also explore how occultation-timing data can best be used to distinguish precessional motion and outline a metric specifically tailored for leveraging and planning occultation observations. As we show, the combination of citizen-science transit observations with occultation observations from more powerful telescopes offers transformational opportunities for a long-term, successful program of transit observations.

\section{Analysis} \label{sec:Analysis}
As discussed in Section \ref{sec:Introduction}, we focus on the Bayesian Information Criterion ${\rm BIC}$ metric for assessing which model most effectively describes a particular timing dataset. We consider what ${\rm BIC}$ values would be expected when timing data exhibiting a specific signal (e.g., precession) are fit with different models. In this context, we use the term ``ephemeris'' to refer to the actual behavior exhibited by the dataset. If a transiting planet exhibits precession, we will describe those data as having a ``precession ephemeris''. We will use the term ``model'' to mean the type of model used to fit those data. For the latter example, we could fit a ``precession model'', which would consider the effects of precession on the timing data, or we could fit a ``linear model'', which would ignore the effects of precession and assume a Keplerian orbit. (Such a model is ``linear'' in the sense that it only includes a term linear in the orbit epoch $E$.) In the following section, we consider what values and variations would be expected for ${\rm BIC}$ when fitting various models to ephemerides.

\subsection{Fitting a Precession Ephemeris with a Precession Model}
We start with a dataset exhibiting precession, i.e., a precession ephemeris, and consider fitting it with a precession model. Of course, in reality, it may not be clear whether a particular dataset exhibits precession, but this combination provides a simple, intuitive starting point. It is worth pointing out that the timing signal for precession may be quite difficult to detect, even for high-precision/long-baseline data, and, moreover, the non-linear dependence on model parameters can mean fitting the model is challenging. In any case, to begin, we assume well-behaved data with a robust and detectable precession signal. We first consider only transit-timing data and will include occultation-timing data in a later section.

Consider a precession ephemeris of mid-transit times $t_i^{\rm tra}$ given by \citep{Gimenez1995}:
\begin{equation}
    t_i^{\rm tra} = T_0 + P_{\rm s} E_i - \left( \frac{e P_{\rm a}}{\pi} \right) \cos \omega_i + \epsilon_i,\label{eqn:transit_precession_ephemeris}
\end{equation}
where $T_0$ is the conjunction time at epoch zero, $E_i$ is the epoch corresponding to observation $i$, $P_{\rm s}$ is the sidereal period (the time between one transit and the next and usually thought of as ``the orbital period''), $e$ is the orbital eccentricity, $\omega_i$ is the argument of pericenter, and $\epsilon_i$ represents the additive noise for point $i$ which has a variance of $\sigma_i^2$. For precession, $\omega_i$ is given by
\begin{equation}
    \omega_i = \omega_0 + \left( \frac{d\omega}{dE} \right) E_i,\label{eqn:omega_evolution}
\end{equation}
where $\omega_0$ is the argument of pericenter at epoch zero and $\frac{d\omega}{dE}$ is the rate of precession. The time between one pericenter passage and the next (which can be longer or shorter than the sidereal period, depending on the sign of the precession) is called the anomalistic period $P_{\rm a}$ and is given in terms of the sidereal period as
\begin{equation}
    P_{\rm a} = \frac{P_{\rm s}}{1 - \frac{1}{2\pi} \frac{d\omega}{dE}}.\label{eqn:anomalistic_period}
\end{equation}

The usual approach to solving for the best-fit model parameters involves minimizing $\chi^2$:
\begin{equation}
\chi^2 = \sum_{i = 0}^{N-1} \sigma_i^{-2} \left( t_i^{\rm tra} - T_0 - P_{\rm s} E_i + \left( \frac{e P_{\rm a}}{\pi} \right) \cos \omega_i \right)^2.
\end{equation} 

With best-fit values for the precession model in hand, we can next evaluate the ${\rm BIC}$ value corresponding to a good model fit. We would expect a good fit for the precession ephemeris to return
\begin{equation}
    {\rm BIC} = \left( N - 5 \right) + 5 \ln N,\label{eqn:BIC_for_precession_model_fit_to_precession_ephemeris}
\end{equation}
where $N$ is the number of data points. Here, we have assumed that $\chi^2$ takes on its expected average value, equal to the number of degrees of freedom, i.e., the number of data points minus the number of model fit parameters \citep{1997iea..book.....T}.

\subsection{Fitting a Linear Model to a Precession Ephemeris}\label{sec:Fitting a Linear Model to a Precession Ephemeris}

Determining via transit-timing data whether a planetary system exhibits Keplerian motion or precessional motion requires comparing a fit using the precession model to a fit using a linear model. For this analysis, we derive a quasi-analytic expression for the difference in ${\rm BIC}$ values between the two models that separates out the dependence on the precession ephemeris parameters, the observed epochs, and the uncertainties. 

Before we work it out, we can ask, ``What is the benefit of such an expression?''. If we have a transit timing dataset, we can easily calculate ${\rm BIC}$ values for each model and determine which is favored (i.e., which has a smaller ${\rm BIC}$). However, as discussed in \citet{2023AJ....166..142J}, the benefit of such an expression is that we can map out a future observing program to maximize the chances for detecting (or corroborating putative) precession signals, given reasonable estimates for the expected timing uncertainties, without taking the observations. Moreover, as we discuss below, this approach also allows us to assess the robustness of a putative detection for precession. 

The $\chi^2$ corresponding to fitting a linear model to a precession ephemeris is given by 
\begin{equation}
    \chi^2 = \sum_{i=0}^{N-1} \sigma_i^{-2} \left( t_i^{\rm tra} - T_0^\prime - P_{\rm s}^\prime E_i \right)^2.\label{eqn:lin_fit_to_prec_ephem_chi_squared}
\end{equation}
Here, $T_0^\prime$ and $P_{\rm s}^\prime$ are the analogous model parameters to the precession model above. However, in the case that the system exhibits precession, their values will not exactly match up to the analogous values in the ephemeris. In other words, the slight curvature from the precession signal will make the values slightly different. How much different? 

We can estimate $T_0^\prime$ and $P_{\rm s}^\prime$ for this case by setting equal to zero the derivatives of $\chi^2$ with respect to each fit parameter, giving the following system of equations:
\begin{equation*}
    \frac{\partial \chi^2}{\partial T_0^\prime} = 0 \Rightarrow 0 = \sum_{i=0}^{N-1} \sigma_i^{-2} \left( t_i^{\rm tra} - T_0^\prime - P_{\rm s}^\prime E_i \right)
\end{equation*}

and

\begin{equation*}
    \frac{\partial \chi^2}{\partial P_{\rm s}^\prime} = 0 \Rightarrow 0 = \sum_{i=0}^{N-1} \sigma_i^{-2} \left( t_i^{\rm tra} - T_0^\prime - P_{\rm s}^\prime E_i \right) E_i.
\end{equation*}

We can re-arrange these equations to arrive at 

\begin{align}
    S_t & = & & S T_0^\prime & + & S_E P_{\rm s}^\prime &\nonumber\\
    S_{Et} & = & & S_E T_0^\prime & + & S_{E^2} P_{\rm s}^\prime &\label{eqn:lin_fit_to_prec_ephem_equation_system}
\end{align}
where we have defined $S_\star$ as a sum related to the variables in the subscript. So, $S \equiv \sum_i \sigma_i^{-2}$, $S_t \equiv \sum_i \sigma_i^{-2} t_i$, $S_{Et} \equiv \sum_i \sigma_i^{-2}E_i t_i$, etc.

Because they have the ephemeris embedded within, the sums involving $t_i^{\rm tra}$ are given by
\begin{align}
    S_t & = & &S T_0& + & S_E P_{\rm s} & - S_{\cos \omega} & \left( \frac{e P_{\rm a}}{\pi} \right)&\nonumber \\
    S_{Et} & = & &S_E T_0& + & S_{E^2} P_{\rm s} & - S_{E \cos \omega} & \left( \frac{e P_{\rm a}}{\pi} \right),&\label{eqn:lin_fit_to_prec_ephem_embedded_sums} 
\end{align}
and we have assumed $\langle \epsilon_i \rangle \approx 0$.

If we define our solution vector as $\vec{b} = \left( T_0^\prime, P_{\rm s}^\prime \right)$, the data vector as $\vec{y} = \left( S_t, S_{Et}\right)$, and the matrix $\mathbf{X}$ built up from the factors attached to $T_0^\prime$, and $P_{\rm s}^\prime$ in Equation \ref{eqn:lin_fit_to_prec_ephem_equation_system}, we can solve for $\vec{b}$ as
\begin{equation*}
    \vec{b} =  \mathbf{X}^{-1} \vec{y}.
\end{equation*}

Solving the equation gives the following:
\begin{align}
    T_0^\prime = T_0 + \bigg[ \frac{S_E S_{E \cos \omega} - S_{E^2} S_{\cos \omega}}{S S_{E^2} - S_E^2} \bigg] \left( \frac{e P_{\rm a}}{\pi} \right) \nonumber\\
    P_{\rm s}^\prime = P_{\rm s} + \bigg[ \frac{S_E S_{\cos \omega} - S S_{E \cos \omega}}{S S_{E^2} - S_E^2} \bigg] \left( \frac{e P_{\rm a}}{\pi} \right). \label{eqn:lin_fit_to_prec_ephem_parameter_corrections}
\end{align}

To simplify the expressions, we will define $\Delta T_0^\prime$ and $\Delta P_{\rm s}^\prime$ to be the terms in the square brackets corresponding to the corrections for each of the fit parameters, respectively.

Now, we can calculate the resulting $\chi^2$:
\begin{equation}
    \chi^2 \approx \left( \frac{e P_{\rm a}}{\pi} \right)^2\ \sum_{i=0}^{N-1} \sigma_i^{-2} \left( \cos \omega_i + E_i \Delta P_{\rm s}^\prime + \Delta T_0^\prime \right)^2 + \sum_{i=0}^{N-1} \sigma_i^{-2} \epsilon_i^2 \approx \left( \frac{e P_{\rm a}}{\pi} \right)^2\ \sum_{i=0}^{N-1} \sigma_i^{-2} \left( \cos \omega_i + E_i \Delta P_{\rm s}^\prime + \Delta T_0^\prime \right)^2 + \left( N - 2 \right),\label{eqn:lin_fit_to_prec_ephem_approximate_chi_squared}
\end{equation}
For convenience when we calculate the variance for $\chi^2$ below, we will use the following definition:
\begin{equation}
    \Sigma \equiv \sum_{i=0}^{N-1} \sigma_i^{-2} \left( \cos \omega_i + E_i \Delta P_{\rm s}^\prime + \Delta T_0^\prime \right)^2.\label{eqn:Sigma_definition}
\end{equation}

Here, we have assumed that squared residuals average out to the number of degrees of freedom, $\left( N - 2 \right)$, plus the correction term involving $\Sigma$. We can use these expressions to calculate the expected average ${\rm BIC}$:
\begin{equation}
    {\rm BIC} = \left( \frac{e P_{\rm a}}{\pi} \right)^2\ \Sigma + \left( N - 2 \right) + 2 \ln N.\label{eqn:BIC_for_lin_fit_to_prec_ephem}
\end{equation}

We can now calculate the $\Delta {\rm BIC}$ for comparing a linear model and a precessional model both fit to a precessional ephemeris. For this calculation, we subtract the ${\rm BIC}$ for the precession model (Equation \ref{eqn:BIC_for_precession_model_fit_to_precession_ephemeris}) from that for the linear model (Equation \ref{eqn:BIC_for_lin_fit_to_prec_ephem}). Written this way, a positive $\Delta {\rm BIC}$ means the data favor the precession over the linear model.
\begin{equation}
    \Delta {\rm BIC} = \left( \frac{e P_{\rm a}}{\pi} \right)^2\ \Sigma - 3 \ln N + 3.\label{eqn:Delta_BIC_lin_fit_vs_prec_fit_to_prec_ephem}
\end{equation}
This expression separates out the dependence on the amplitude of the precession signal ($e P_{\rm a}/\pi$). Not surprisingly, the larger the amplitude, the more quickly $\Delta {\rm BIC}$ grows and favors precession over no-precession, all other things being equal. However, the negative term means that a sufficient number of data points are required for $\Delta {\rm BIC}$ to exceed zero and for the data to favor precession.

\begin{figure}
    \centering
    \includegraphics[width=\linewidth]{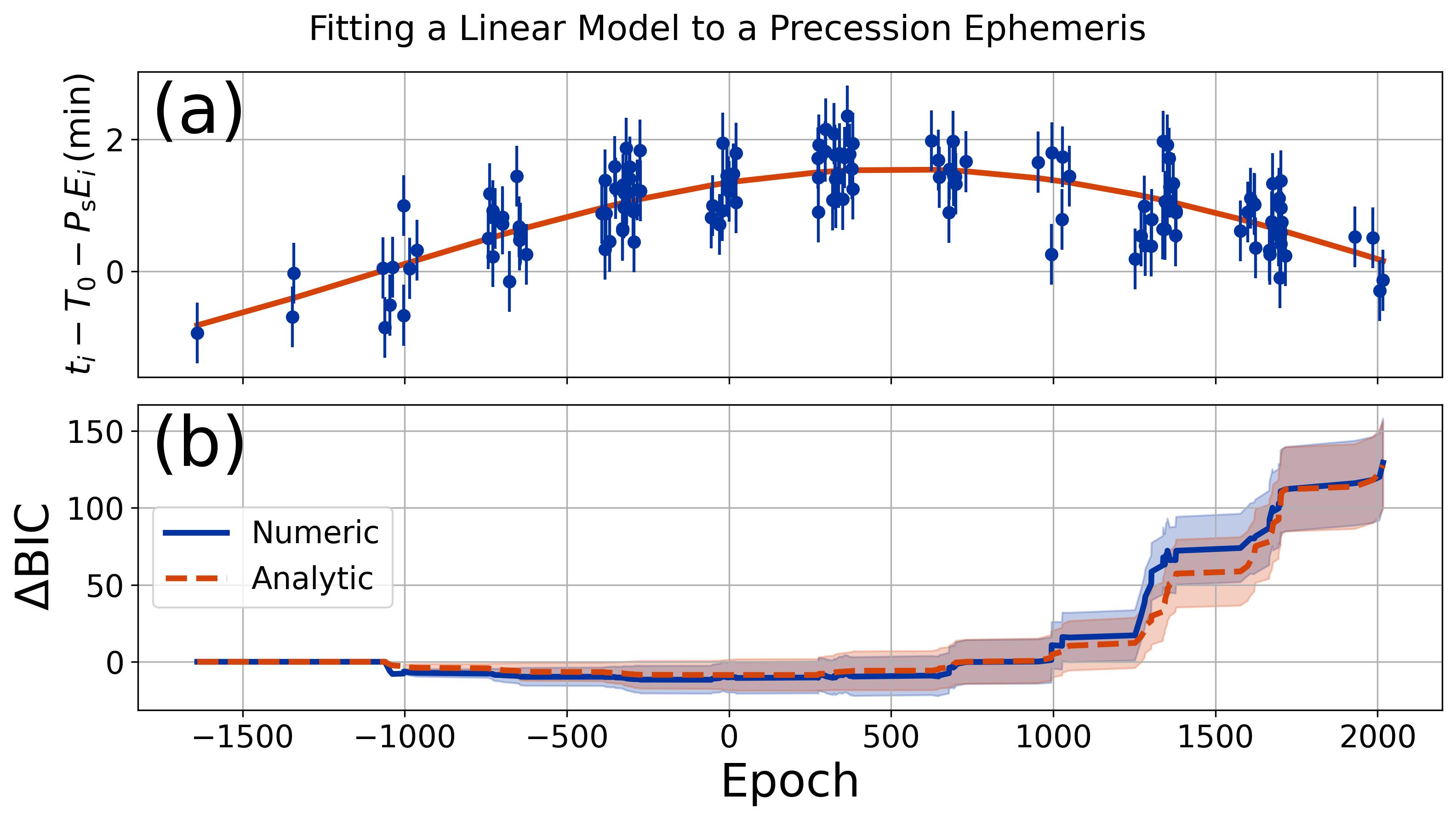}
    \caption{(a) The blue dots represent synthetic transit-timing data based on (but not actually using) data reported in \citet{2020ApJ...888L...5Y}. For this synthetic dataset, we took the orbital epochs of transit observations of WASP-12 b and generated a synthetic precession ephemeris according to Equation \ref{eqn:transit_precession_ephemeris}. Into that dataset, we injected Gaussian noise with a timing scatter equal to the average timing scatter for the data reported in \citet{2020ApJ...888L...5Y}. The precession ephemeris parameters used were $T_0 =2456305.45488\,{\rm BJD}$, $P_{\rm s} = 1.091419633\,{\rm days}$, $e = 0.00310$, $d\omega/dE = 0.000984\,{\rm rad\ orb^{-1}}$, and $\omega_0 = 2.62\,{\rm rad}$. The orange line shows the best-fit precession model for the whole dataset. (b) Evolution of $\Delta {\rm BIC}$ comparing a linear and a precession model fit to a precession ephemeris. $\Delta {\rm BIC} > 0$ favors the precession over the linear model. The solid, blue line shows the direct numeric estimate of $\Delta {\rm BIC}$, while the dashed, orange line shows implementation of the quasi-analytic estimate, Equation \ref{eqn:Delta_BIC_lin_fit_vs_prec_fit_to_prec_ephem}. The shaded blue and orange regions show the range of $\Delta {\rm BIC}$ values expected from random variations.}
    \label{fig:Testing_Linear_Fit_to_Precession_Ephemeris}
\end{figure}

Figure \ref{fig:Testing_Linear_Fit_to_Precession_Ephemeris} shows an example application of Equation \ref{eqn:Delta_BIC_lin_fit_vs_prec_fit_to_prec_ephem} to synthetic data. For this example, we assumed the best-fit precession model for the transit-timing data reported for WASP-12 b in \citet{2020ApJ...888L...5Y}. (N.B., WASP-12 b exhibits tidal decay and not precession.) Panel (a) shows those data with the linear portion of the ephemeris subtracted out (blue dots), along with the best-fit precession model, while panel (b) shows how $\Delta {\rm BIC}$ evolves as time passes and the data mount, comparing the direct numeric to the quasi-analytic estimate (Equation \ref{eqn:Delta_BIC_lin_fit_vs_prec_fit_to_prec_ephem}). The two lines track very closely, and both remain below zero until just past $E = 500$, indicating the data favor the linear over the precession model until that point (spanning over 2000 total epochs).

However, the two curves do not precisely agree along the whole track. Why not? Recall that Equation \ref{eqn:Delta_BIC_lin_fit_vs_prec_fit_to_prec_ephem} assumes the $\chi^2$ value for fitting a linear model and the $\chi^2$ value for fitting a precession model to the data both take on their average values. For any particular dataset, though, $\chi^2$ is unlikely to exactly equal the average value. Instead, we expect $\chi^2$ to lie within some range of the average value. In other words, we need to consider both the average value and the variance for $\Delta{\rm BIC}$. The variance for $\Delta{\rm BIC}$ is given by $\sigma^2_{\Delta {\rm BIC}} = \sigma^2_{\chi^2_{\rm prec}} + \sigma^2_{\chi^2_{\rm lin}} - 2\ {\rm covar}\left( \chi^2_{\rm prec}, \chi^2_{\rm lin}\right)$, where $\sigma^2_{\chi^2_{\rm prec}}$ is the variance for $\chi^2_{\rm prec}$ that arises from fitting the precession ephemeris with a precession model, $\sigma^2_{\chi^2_{\rm lin}}$ is the variance for $\chi^2_{\rm lin}$ for fitting the precession ephemeris with a linear model, and ${\rm covar}\left( \right)$ is the covariance between them.

For fitting a precession ephemeris with a precession model, we have $\sigma^2_{\chi^2_{\rm prec}} = 2 \left( N - 5\right)$, i.e., twice the number of degrees of freedom, since the relevant distribution is the standard $\chi^2$ distribution. However, for fitting a precession ephemeris with a linear model, we have a non-centered $\chi^2$ distribution, which is traditionally described with a non-centrality parameter $\lambda$ representing the offset from $\chi = 0$ \citep{Johnson1970}. Equation \ref{eqn:Sigma_definition} gives the relevant non-centrality parameter, $\lambda = \left( e P_{\rm a}/\pi \right)^2 \Sigma$. The non-central $\chi^2$ distribution has a variance that depends on the number of degrees of freedom and the non-centrality parameter as $\sigma^2_{\chi^2_{\rm lin}}= 2 \left( N - 2 \right) + 4 \left( e P_{\rm a}/\pi \right)^2 \Sigma$.

We turn next to the covariance term. If the $\chi^2$ values arising from the linear and precession models were statistically independent, then the covariance between them would be zero \citep{Johnson1970}. Since both fits involve the same data, the $\chi^2$ values are not independent, and the covariance is decidedly non-zero. However, for the small eccentricities relevant to our problem (a few percent or less), the linear terms in the ephemeris dominate, meaning the two $\chi^2$ values move in tandem, and the covariance is positive. Because the expression for the variance of $\Delta {\rm BIC}$ involves subtracting that covariance (the negative sign arises from the definition of $\Delta {\rm BIC}$), a positive covariance can only reduce the variance for $\Delta {\rm BIC}$. Numerical experimentation with representative uncertainties and model parameters shows the magnitude of twice the covariance is typically less than 30\% of total variance for $\Delta {\rm BIC}$. For larger eccentricities, the impact of the $\cos$ term in the ephemeris grows, and for very large eccentricities ($e \rightarrow 1$), the covariance can flip sign. However, by that point, $\sigma^2_{\chi^2_{\rm lin}}$ grows so large (i.e., the linear model is a very poor fit) that twice the covariance is negligible, whatever its sign. Thus, we neglect the covariance term and approximate the variance on $\Delta {\rm BIC}$ as 

\begin{equation}
\sigma^2_{\Delta {\rm BIC}} \approx \sigma^2_{\chi^2_{\rm prec}} + \sigma^2_{\chi^2_{\rm lin}} = 2 \left( N - 7 \right) + 4 \left( \frac{eP_{\rm a}}{\pi} \right)^2 \Sigma,\label{eqn:approx_sigma_squared_Delta_BIC_linear_vs_precession_model_to_precession_ephemeris}
\end{equation}
This approach should over-estimate the variance and thereby provide a conservative estimate of uncertainty. In the following sections, we use the same approximation since the same logic applies for all combinations of models and ephemerides considered here. 

As we discuss below, $\Delta {\rm BIC}$ should exceed zero by at least $\sigma^2_{\Delta {\rm BIC}}$ to conclude that a dataset favors non-Keplerian motion. Including only the first term on the right-hand side of Equation \ref{eqn:approx_sigma_squared_Delta_BIC_linear_vs_precession_model_to_precession_ephemeris} would also provide a useful (and simpler) threshold for such an assessment -- of course, $\Delta {\rm BIC}$ must at least exceed that latter threshold to satisfy Equation \ref{eqn:approx_sigma_squared_Delta_BIC_linear_vs_precession_model_to_precession_ephemeris}. We will return to this point later. The shaded orange region in Figure \ref{fig:Testing_Linear_Fit_to_Precession_Ephemeris} shows $\pm \sqrt{ \sigma^2_{\Delta {\rm BIC}} }$, corresponding to a 68\% probability for ${\Delta {\rm BIC}}$ to lie within that range, which the blue line does.

This estimate of the plausible range for $\Delta {\rm BIC}$ also provides a means for assessing the robustness of a putative detection of precession in dataset. For instance, in Figure \ref{fig:Testing_Linear_Fit_to_Precession_Ephemeris}, the blue line grows somewhat faster than the orange line after $E = 1000$. This result indicates that random scatter in the data gives rise to a mild misestimate for $\Delta {\rm BIC}$, meaning the wrong model might be favored. Indeed, the shaded orange band only ceases to encompass $\Delta {\rm BIC} = 0$ beyond $E = 1250$ (about 3000 total epochs of observational baseline). 

\subsection{Fitting a Quadratic Model to a Precession Ephemeris}\label{sec:Fitting_a_Quadratic_Model_to_a_Precession_Ephemeris}
Next, we explore the case of fitting a precession ephemeris with a quadratic model. This case might arise as curvature in a timing dataset starts to manifest but it is not yet clear whether to favor tidal decay or precession. The precession ephemeris $t_i^{\rm tra}$ for transit mid-times is given in Equation \ref{eqn:transit_precession_ephemeris}. To fit a quadratic model to this ephemeris, we construct our $\chi^2$ expression as
\begin{equation}
    \chi^2 = \sum_{i=0}^{N-1} \sigma_i^{-2} \left( t_i^{\rm tra} - T_0^\prime - P_{\rm s}^\prime E_i - \left( \frac{1}{2} \frac{dP}{dE} \right) E_i^2 \right)^2.\label{eqn:prec_ephem_chi_squared}
\end{equation}

We can employ the same linear regression approach as in Section \ref{sec:Fitting a Linear Model to a Precession Ephemeris}, which we walk through in Appendix \ref{sec:Appendix_Fitting_Quadratic_Model_to_Precession_Ephemeris}. With the resulting expressions in hand, we can calculate the $\chi^2$ that results from fitting the quadratic model to the precession ephemeris:
\begin{equation}
    \chi^2 = \left( N - 3 \right) + \left( \frac{e P_{\rm a}}{\pi} \right)^2 \sum_{i = 0}^{N-1} \sigma_i^{-2} \left( \cos \omega_i + \Delta T_0^\prime + \Delta P_{\rm s}^\prime E_i + \Delta \left( \frac{dP}{dE} \right) E_i^2 \right)^2,\label{eqn:quad_fit_to_precession_chi_squared}
\end{equation}
where the $\Delta *$ terms are given in Appendix \ref{sec:Appendix_Fitting_Quadratic_Model_to_Precession_Ephemeris}.

With Equation \ref{eqn:quad_fit_to_precession_chi_squared}, we can construct the BIC for fitting a precession ephemeris with a quadratic model:
\begin{equation}
    {\rm BIC} = \left( N - 3 \right) + \left( \frac{e P_{\rm a}}{\pi} \right)^2 \sum_{i = 0}^{N-1} \sigma_i^{-2} \left( \cos \omega_i + \Delta T_0^\prime + \Delta P_{\rm s}^\prime E_i + \Delta \left( \frac{dP}{dE} \right) E_i^2 \right)^2 + 3 \ln N. \label{eqn:BIC_fit_quad_to_prec}
\end{equation}

Then we can estimate $\Delta{\rm BIC}$ for comparing this fit to fitting a precession model to a precession ephemeris:
\begin{equation}
    \Delta{\rm BIC} = \left( \frac{e P_{\rm a}}{\pi} \right)^2 \sum_{i = 0}^{N-1} \sigma_i^{-2} \left( \cos \omega_i + \Delta T_0^\prime + \Delta P_{\rm s}^\prime E_i + \Delta \left( \frac{dP}{dE} \right) E_i^2 \right)^2 - 2 \ln N + 2.\label{eqn:Delta_BIC_quad_to_prec}
\end{equation}
Cast in this way, $\Delta{\rm BIC} > 0$ implies that the data favor a precession model over a quadratic. We can also estimate the corresponding uncertainty 
\begin{equation}
\sigma^2_{\Delta {\rm BIC}} \approx 2 \left( N - 8\right) + 4 \left( \frac{e P_{\rm a}}{\pi} \right)^2 \Sigma,\label{eqn:sigma_squared_Delta_BIC_quad_vs_precession_model_to_precession_ephemeris} 
\end{equation}
where $\Sigma$ is the summation term in Equation \ref{eqn:Delta_BIC_quad_to_prec}, analogous to the expression from Section \ref{sec:Fitting a Linear Model to a Precession Ephemeris}.

\begin{figure}
    \centering
    \includegraphics[width=\linewidth]{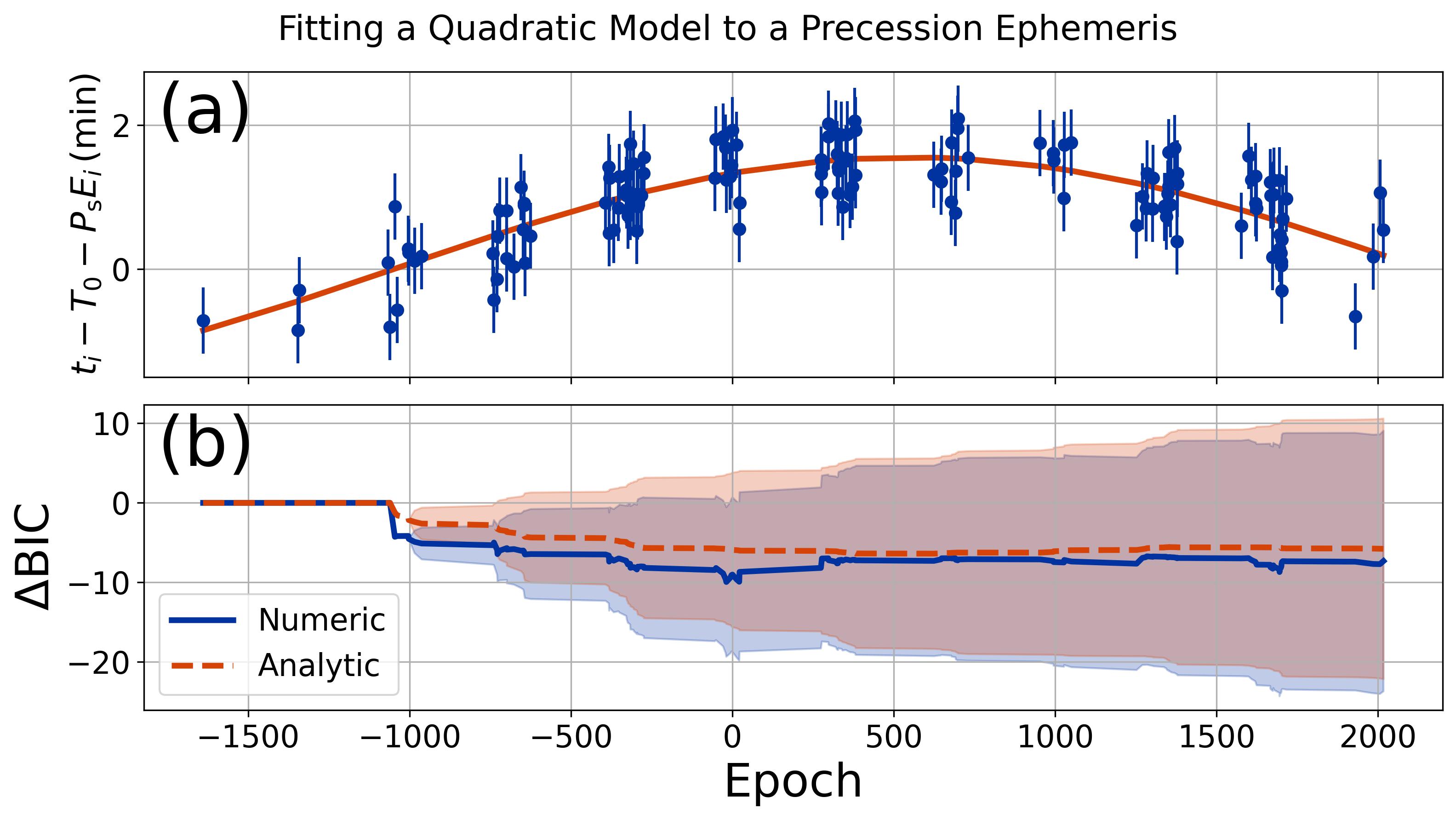}
    \caption{(a) The blue dots represent a synthetic transit-timing data based on the same approach as used in Figure \ref{fig:Testing_Linear_Fit_to_Precession_Ephemeris}. The symbols have the same meaning as in that figure. (b) Evolution of $\Delta {\rm BIC}$ comparing a quadratic and a precession model, both fit to a precession ephemeris. 
    $\Delta {\rm BIC} > 0$ favors the precession over the quadratic model. The solid blue line shows the direct numeric estimate of $\Delta {\rm BIC}$, while the dashed orange line shows implementation of the quasi-analytic estimate, Equation \ref{eqn:Delta_BIC_quad_to_prec}. The shaded blue and orange regions show the range of $\Delta {\rm BIC}$ values expected from random variations.}
    \label{fig:Testing_Quad_Fit_to_Precession_Ephemeris}
\end{figure}

Figure \ref{fig:Testing_Quad_Fit_to_Precession_Ephemeris} shows the analogous situation to Figure \ref{fig:Testing_Linear_Fit_to_Precession_Ephemeris}, where, instead of a linear model fit to the precession ephemeris, we have used a quadratic model. Unlike Figure \ref{fig:Testing_Linear_Fit_to_Precession_Ephemeris}, however, we see that neither the numerical nor quasi-analytic $\Delta{\rm BIC}$ exceed zero for the whole span of observations. The obvious conclusion is that, although the O-C plot shows clear curvature, this synthetic dataset would not suffice to distinguish between precession or tidal decay as the cause. To yield a positive $\Delta{\rm BIC}$ faster, different parameters would be needed: a larger eccentricity, a faster precession rate, or smaller timing uncertainties. 

\subsection{Fitting a Precession Model to a Quadratic Ephemeris}\label{sec:Fitting a Precession Model to a Quadratic Ephemeris}
To fit a precession model to a quadratic ephemeris, we construct our $\chi^2$ expression as
\begin{equation}
    \chi^2 = \sum_{i=0}^{N-1} \sigma_i^{-2} \left( t_i^{\rm tra} - T_0^\prime - P_{\rm s}^\prime E_i + \left( \frac{e P_{\rm a}}{\pi} \right) \cos \left( \frac{d\omega}{dE} E_i + \omega_0 \right) \right)^2.\label{eqn:quad_ephem_chi_squared}
\end{equation}
For this problem, our fit parameters are $T_0^\prime$, $P_{\rm s}^\prime$, $e$, $d\omega/dE$, and $\omega_0$ (recall that $P_{\rm s}$ and $d\omega/dE$ determine $P_{\rm a}$ -- Equation \ref{eqn:anomalistic_period}). 

With the dependence on $\cos \omega_i$, this problem is not amenable to linear regression and therefore we cannot construct a quasi-analytic expression for $\Delta {\rm BIC}$ analogous to Equations \ref{eqn:Delta_BIC_lin_fit_vs_prec_fit_to_prec_ephem} and \ref{eqn:Delta_BIC_quad_to_prec}. However, we can still explore the evolution of $\Delta {\rm BIC}$ to see how evidence favoring the quadratic model (tidal decay) over the precession model grew as timing data for WASP-12 b mounted. A best-fit precession model can also be propagated forward in time beyond the end of currently available observational data for a particular ephemeris to predict how $\Delta {\rm BIC}$ would be expected to evolve for a proposed set of future observations -- more on this point below. We can also estimate the expected variance:
\begin{equation}
    \sigma^2_{\Delta {\rm BIC}} > 2 \left( N - 8\right).\label{eqn:sigma_squared_Delta_BIC_precession_vs_quad_model_to_quad_ephemeris}
\end{equation}

\begin{figure}
    \centering
    \includegraphics[width=\linewidth]{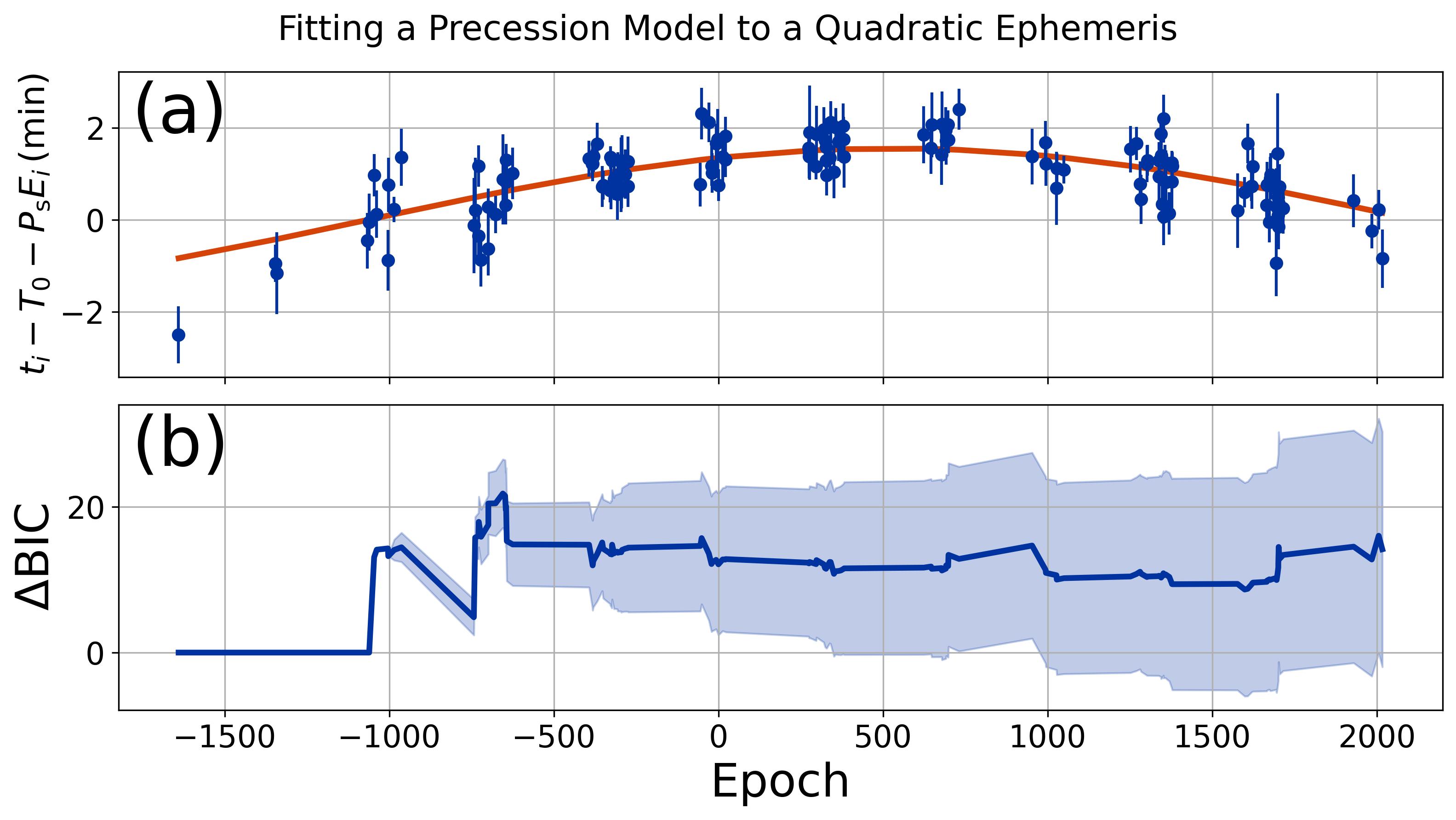}
    \caption{(a) The blue dots show the real transit timing data from \citet{2020ApJ...888L...5Y} for WASP-12 b. The orange line shows a precession model fit to those data. (b) Evolution of $\Delta {\rm BIC}$ comparing a precession and a quadratic model, both fit to the real data in panel (a); $\Delta {\rm BIC}>0$ indicates a preference for a quadratic model. On the basis of these transit-timing, occultation timing, as well as radial velocity data, \citet{2020ApJ...888L...5Y} concluded that the transit-timing variations seen for WASP-12 b result from tidal decay, so we consider the example shown here to represent a quadratic ephemeris. No closed-form quasi-analytic expression is available to estimate $\Delta {\rm BIC}$ for fitting a precession model to a quadratic ephemeris, so panel (b) only includes what we have called in previous figures the ``Numeric'' estimate, along with the estimated range of variation (shaded blue).}
    \label{fig:Testing_Precession_Fit_to_Quad_Ephemeris}
\end{figure}

Since we do \added{not} have an analytic expression for $\chi^2$ for this instance of $\Delta {\rm BIC}$, we cannot write a closed form expression for the non-centrality parameter as we did for the other instances. Instead, we take Equation \ref{eqn:sigma_squared_Delta_BIC_precession_vs_quad_model_to_quad_ephemeris} as the minimum threshold required for $\Delta {\rm BIC}$ as discussed above in Section \ref{sec:Fitting a Linear Model to a Precession Ephemeris}. The results here provide an interesting point of comparison to the results in Figure \ref{fig:Testing_Quad_Fit_to_Precession_Ephemeris} in which we considered fitting a quadratic model to a precession ephemeris. In that figure, $\Delta {\rm BIC}$ remained in negative territory throughout the whole dataset but with relatively large error bars, meaning those (synthetic) data were not able to robustly distinguish whether a quadratic or a precession model was preferred. For the present case shown in Figure \ref{fig:Testing_Precession_Fit_to_Quad_Ephemeris}, $\Delta {\rm BIC}$ only just \added{begins to exceed} zero by the above simplified estimate for $\sigma_{\Delta {\rm BIC}}$ by the last data point, but we have already said Equation \ref{eqn:sigma_squared_Delta_BIC_precession_vs_quad_model_to_quad_ephemeris} underestimates the uncertainty on $\Delta {\rm BIC}$. A reasonable conclusion therefore is that timing data from transits alone do not suffice to decide whether tidal decay or precession is responsible for inducing transit-timing variations. Indeed, as discussed in \citet{2020ApJ...888L...5Y}, including occultation timing data was key to attributing WASP-12 b's timing variations to tidal decay and not to precession. 

It is worth noting that the difficulty of distinguishing between tidal decay and precession stands in contrast to distinguishing between a Keplerian or non-Keplerian orbit for a planet. As discussed in \citet{2020ApJ...888L...5Y} and \citet{2023AJ....166..142J}, the $\Delta {\rm BIC}$ value for comparing a linear and a quadratic model using all transit-timing data for WASP-12 b exceeds 200, robustly inconsistent with a Keplerian orbit. The key point here is that with transit-timing data alone it may be ambiguous which non-Keplerian effect is in play, even when we can definitively detect such an effect. Below, we consider the role of occultation-timing data in this context.

\subsection{Fitting a Linear Model to a Quadratic Ephemeris}
Although our focus here is on precession, we also briefly address the case of fitting a linear model to a quadratic ephemeris. This case applies when a system exhibits tidal decay and we are trying to assess whether the timing data favor decay over Keplerian motion. Many of the details for this case are discussed in \citet{2023AJ....166..142J}, so we only report the results here.

Analogous to our prior results in the current study, we have the following expression for comparing a linear to a quadratic model in the case of a quadratic ephemeris:
\begin{equation}
    \Delta {\rm BIC} = \frac{1}{4} \left( \frac{dP}{dE} \right)^2 \sum_{i=0}^{N-1} \sigma_i^{-2} \left( E_i^2 - \Delta P^\prime E_i - \Delta T_0^\prime \right)^2 - \ln N + 1,\label{eqn:Delta_BIC_lin_to_quad}
\end{equation}
where $\frac{dP}{dE}$ is the change in orbital period from one orbit to the next (i.e., the decay rate) and $\Delta P^\prime$ and $\Delta T_0^\prime$ are the corrections to the orbital period and conjunction time analogous to the values in the previous sections -- \citet{2023AJ....166..142J} gives the closed form expression for those terms. We also have the analogous variance on $\Delta {\rm BIC}$:

\begin{equation}
    \sigma_{\Delta {\rm BIC}}^2 \approx 2 \left( N - 5 \right) + 4\times \frac{1}{4} \left( \frac{dP}{dE} \right)^2 \Sigma.
\end{equation}\label{eqn:sigma_Delta_BIC_lin_to_quad}
We use these expressions in Section \ref{sec:Application} below.

\subsection{Including Occultation Timing Data}

Next, we consider how occultation timing data impact inferences regarding precession. Of course, the smaller depths for occultations as compared to transits mean the corresponding timing data are less precise, as discussed in \citet{2023AJ....166..142J}. For instance, the transits of WASP-12 b discussed in \citet{2020ApJ...888L...5Y} have depths of about 2\% and yield timing precisions of about $40\,{\rm s}$, while the occultations have depths of about 0.5\% and timing precisions of about $70\,{\rm s}$. Since occultation depths are dependent on, for example, variable temperature and atmospheric properties of a planet while the transit depth depends solely on the planetary and stellar radius values, the expected depths and timing precisions for occultations of different planets with similar transit depths may vary widely.

The linear ephemeris for occultations is given by 
\begin{equation}
    t_i^{\rm occ} = T_0 + P_{\rm s} \left( E_i + \frac{1}{2} \right).\label{eqn:occultation_linear_ephemeris}
\end{equation}

The quadratic ephemeris is given by
\begin{equation}
    t_i^{\rm occ} = T_0 + P_{\rm s} \left( E_i + \frac{1}{2} \right) + \left( \frac{1}{2} \frac{dP}{dE} \right) E_i^2.\label{eqn:occultation_quadratic_ephemeris}
\end{equation}

The precession ephemeris is given by
\begin{equation}
    t_i^{\rm occ} = T_0 + P_{\rm s} E_i + \frac{P_{\rm a}}{2} + \left(\frac{e P_{\rm a}}{\pi} \right) \cos \left( \frac{d\omega}{dE} E_i + \omega_0 \right),\label{eqn:occultation_precession_ephemeris}
\end{equation}
where the variables have all the same meanings as in the analogous equations for the transit ephemeris. The key differences are changes to $E_i + \frac{1}{2}$ in the $P_{\rm s}$ terms and a plus sign instead of a minus sign appending the $\cos$ term in the precession ephemeris. In fact, this change in sign will be important for distinguishing tidal decay and precession.

In principle, we can derive $\Delta {\rm BIC}$ expressions that consider occultations analogous to the expressions given in previous sections. However, transit-timing data are usually much more plentiful than occultation-timing data, and occultation-timing data have much larger uncertainties. Consequently, the latter data have much less utility for distinguishing Keplerian from non-Keplerian orbits (i.e., linear from either quadratic or precession models). As we discuss below, though, they are useful for distinguishing between non-linear models, e.g., tidal decay from precession.

\begin{figure}
    \centering
    \includegraphics[width=\linewidth]{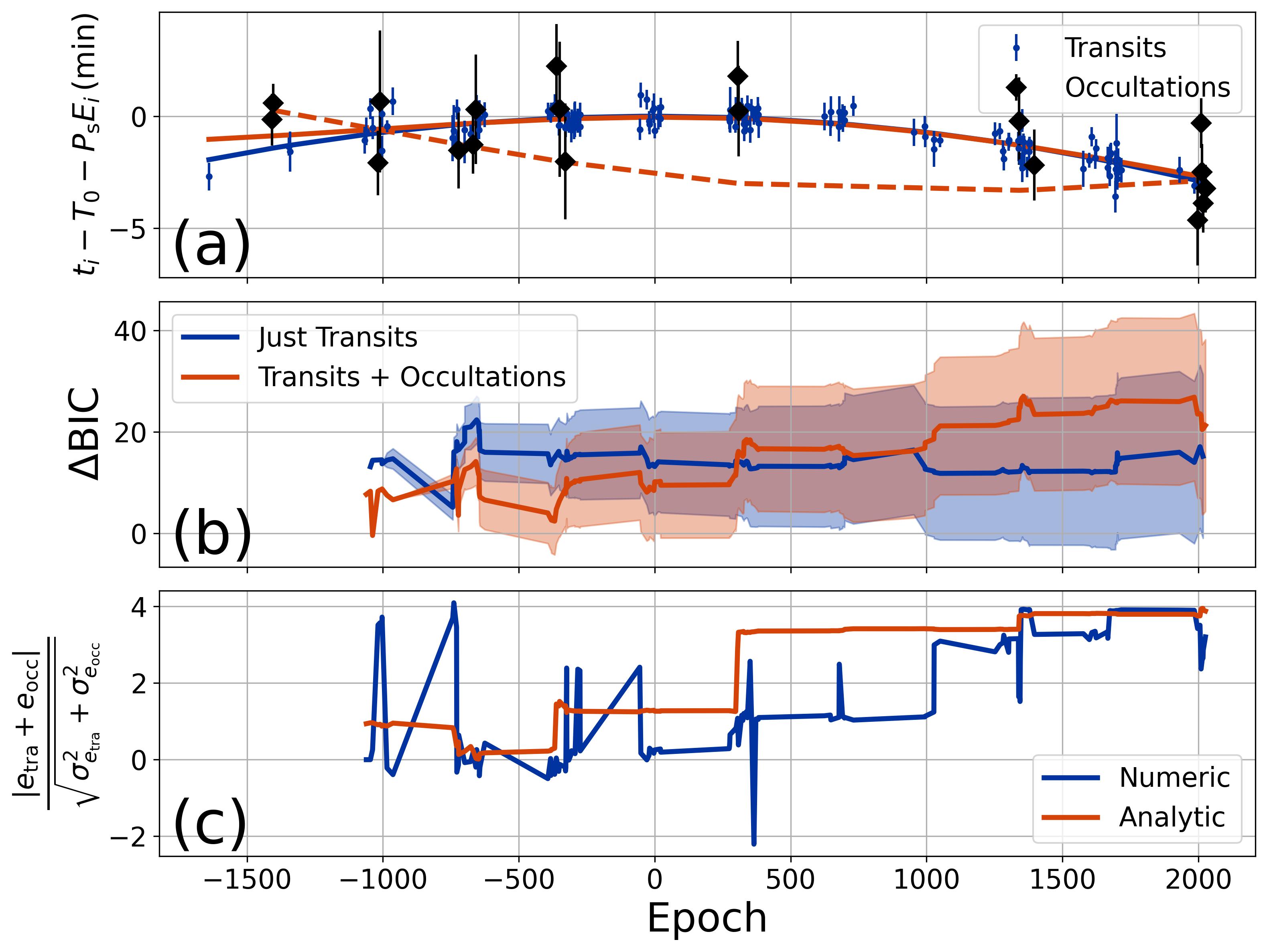}
    \caption{Exploring the impact of occultation timing data on model-fitting. (a) The blue dots show the transit-timing data from Yee et al. (2020) for WASP-12 b, while the black diamonds show the occultation-timing data. The solid, blue line shows a quadratic (tidal decay) model fit to both datasets, while the solid, orange line shows a transit-timing precession model fit to both datasets. The dashed, orange line shows the model occultation times for the same precession parameters. Because the occultation times have much larger error bars, the best-fit model tolerates a larger disagreement between the observed and the modeled occultation times. (b) The evolution of $\Delta{\rm BIC}$ for comparing the quadratic and the precession model both fit to the mounting data. The blue line shows the the evolution considering only the transit data (the blue shaded region shows the expected range), while the orange line shows the evolution considering both the transit and occultation data (again, the orange shaded region shows the range). $\Delta{\rm BIC} > 0$ favors tidal decay over precession. (c) Evolution of the eccentricity metric (Equation \ref{eqn:eccentricity_metric}). The blue line shows the direct, numeric result, while the orange line shows the approximate quasi-analytic approach (Equation \ref{eqn:anal_eccentricity}).}
    \label{fig:Adding_Occultation_Timing_Data}
\end{figure}

To show the impact of occultation data, we turn again to the WASP-12 b dataset reported in \citet{2020ApJ...888L...5Y}, this time including the occultation timing data as well as the transit timing data. For WASP-12 b, considerable ink was spilt trying to argue over whether the decisively non-Keplerian transit-timing data favored tidal decay or precession \citep{2019MNRAS.482.1872B, 2017AJ....153...78C, 2011ApJ...727..125C, 2019MNRAS.490.1294B}. Using those data, panel (a) in Figure \ref{fig:Adding_Occultation_Timing_Data} shows a quadratic model (solid blue line) fit to both the transit and occultation timing data. We also show a precession model fit to the transit and occultation timing data (solid orange line). The occultation data clearly line up alongside the transit data, consistent with the conclusion of that WASP-12 exhibits tidal decay. 

The $\Delta {\rm BIC}$ values shown in panel (b) represent a comparison of a quadratic vs. a precession model fit to the data -- here, $\Delta {\rm BIC} > 0$ means the data indicate a preference for the quadratic model. The blue line shows how $\Delta {\rm BIC}$ evolves, considering only transit-timing data, while the orange line also includes the occultation-timing data. Not surprisingly, the final $\Delta {\rm BIC}$ values at $E = 2000$ both favor the quadratic to the precession model, but interestingly, the values consistently exceed zero, with expected ranges (shown by the shaded regions) very often inconsistent with zero. In other words, the WASP-12 b timing data, whether including occultations or not, have favored tidal decay over precession for much of the observational baseline for many years. Also interesting, though, is the fact that the final $\Delta {\rm BIC}$ value considering occultation-timing data (orange) is 21, while the final $\Delta {\rm BIC}$ value ignoring occultations (blue) is 15 -- a difference of only 6 -- and each $\Delta {\rm BIC}$ value has an uncertainty of $\sigma_{\Delta {\rm BIC}} > 16$ (Equation \ref{eqn:sigma_squared_Delta_BIC_precession_vs_quad_model_to_quad_ephemeris}), meaning the occultation-including $\Delta {\rm BIC}$ is only marginally inconsistent with zero. 

The key distinguishing characteristic between decay and precession was the downward concavity of the occultation-timing data -- the dashed orange line in Figure \ref{fig:Adding_Occultation_Timing_Data} shows what the occultation-timing would have looked like if the system exhibited precession instead of decay. Right around $E = 250$ occurs the maximum discrepancy between the actual data and the expected signal for precession. The orange $\Delta {\rm BIC}$ does show a modest uptick at that point, but the increase is not qualitatively impressive and lies well within the expected variations. However, we can formulate a metric specially tailored to distinguish decay from precession by leveraging the difference in the sign for the cosine terms for the precession ephemeris, as seen in comparing Equation \ref{eqn:transit_precession_ephemeris} to \ref{eqn:occultation_precession_ephemeris}. 

For that metric, we fit the precession model (with an assumed minus sign in front of the cosine term) to only the transit data but without restricting the sign of the cosine term. In other words, we allow the eccentricity to be positive or negative and allow the transit-timing data to determine the sign of the eccentricity, defining the resulting eccentricity as $e_{\rm tra}$. Likewise, we can define $e_{\rm occ}$ resulting from fitting only the occultation-timing data (again, with an assumed minus sign in front of the cosine term). In the case that the timing data support precession, we expect $e_{\rm tra} = -e_{\rm occ}$. In the case that the timing data support tidal decay, we expect $e_{\rm tra} = e_{\rm occ} = e$. Thus, the expression $e_{\rm tra} + e_{\rm occ}$ statistically consistent with zero comports with precession, while $e_{\rm tra} + e_{\rm occ}$ statistically inconsistent with zero favors tidal decay. If we have uncertainties associated with $e_{\rm tra}$ and $e_{\rm occ}$, $\sigma_{e_{\rm tra}}$ and $\sigma_{e_{\rm occ}}$, respectively, we can cast this expectation for tidal decay as 
\begin{equation}
    \frac{| e_{\rm tra} + e_{\rm occ} |}{\sqrt{\sigma_{e_{\rm tra}}^2 + \sigma_{e_{\rm occ}}^2}} > H,\label{eqn:eccentricity_metric}
\end{equation}
where we have added the uncertainties in quadrature and $H$ is the desired threshold for deciding the data favor tidal decay. 

The blue curve in Figure \ref{fig:Adding_Occultation_Timing_Data}(c) shows the evolution of this metric as the real timing data for WASP-12 b mounted. For this calculation, we fixed $T_0$ and $P_{\rm s}$ at the best-fit values that arose from fitting all the available data (whether transit-timing or occultation-timing) and allowed the other precession model parameters ($d\omega/dE$, and $\omega_0$) to float. To fit those latter parameters, we used the Trust Region Reflective algorithm \citep{1982SJNA...19..409S} as implemented in scipy curve\_fit \citep{2020SciPy-NMeth}. It is worth noting that trying to fit $T_0$ and $P_{\rm s}$ as well as all the precession parameters at once often resulted in poor fits, depending on the initial guesses we chose. We found, instead, that first fitting $T_0$ and $P_{\rm s}$ and then fitting the precession parameters with $T_0$ and $P_{\rm s}$ fixed at those best-fit values tended to result in convergent solutions. This behavior is not surprising since the model is vastly more sensitive to $T_0$ and $P_{\rm s}$ than the precession parameters.

Looking at Figure \ref{fig:Adding_Occultation_Timing_Data}(c), we see that the numeric metric initially shows a preference for tidal decay over precession; however, the large error bars on occultation-timing data around $E = -750$ suddenly drives the eccentricity metric to near zero, suggesting the data no longer favor tidal decay over precession, and the strong evidence for decay does not emerge again in the blue curve until $E > 1000$, when more occultation-timing data were collected. By the end, the metric clearly exceeds 3, suggesting strong evidence for tidal decay over precession.

The orange curve in Figure \ref{fig:Adding_Occultation_Timing_Data}(c) shows a slightly different, quasi-analytic approach to estimating the eccentricity metric. For that curve, we first subtracted the linear portion of the best-fit precession model ($T_0 + P_{\rm s} E_i$) from the data and then divided that residual by $P_{\rm a}/\pi$, again, determined by using all the available data. In principle, these operations leave a scaled residual $\Delta t_i^\prime$ that exhibits only the departure from a linear ephemeris. Finally, we used the function $\Delta t_i^\prime = e \cos \omega_i$ to fit only the $e$ value to the residual (fixing all other precession parameters at their best-fit values). This approach gives the following expression for the transit-/occultation-associated eccentricity 
\begin{equation}
    e_{\rm tra/occ} = -\frac{\sum_{i \in {\rm tra/occ}} \sigma_{\Delta t^\prime_i}^{-2} \Delta t^\prime_i \cos\omega_i }{\sum_{i \in {\rm tra/occ}} \sigma_{\Delta t^\prime_i}^{-2} \cos^2\omega_i } = -\frac{S_{\Delta t^\prime \cos \omega}}{S_{\cos^2\omega}},\label{eqn:anal_eccentricity}
\end{equation}
where $\sigma_{\Delta t^\prime_i} \approx \pi \sigma_i/P_{\rm a}$, i.e., the timing uncertainties divided by $P_{\rm a}/\pi$, and we have again used the $S_*$ notation to indicate implied summations. (We have assumed the uncertainties on individual timing data points dominate over uncertainties on other ephemeris parameters, $T_0$ and $P_{\rm s}$.) We can also estimate an associated uncertainty for $e_{\rm tra/occ}$:

\begin{equation}
    \sigma_{e_{\rm tra/occ}} \approx \left(  \sum_{i \in {\rm tra/occ}} \left( \frac{\partial e_{\rm tra/occ}}{\partial \left( \Delta t^\prime_i \right)} \right)^2 \sigma_{\Delta t_i^\prime}^2 \right)^{1/2} = S_{\cos^2 \omega}^{-1/2}.\label{eqn:sigma_anal_eccentricity}
\end{equation}
Then, the corresponding eccentricity metric can be formed using these equations. As discussed in previous sections, the primary benefit of these quasi-analytic expressions is to extrapolate observations forward in time for a system for which precession with estimated $d\omega/dE$ and $\omega_0$ values is suspected and to check what observing program would best bolster evidence for that precession. As Figure \ref{fig:Adding_Occultation_Timing_Data}(c) shows, the eccentricity metric estimated using this quasi-analytic approach somewhat overestimates the metric along the evolutionary track, but by the end of the observational sequence the numeric and quasi-analytic estimates match very closely. 

\begin{figure}
    \centering
    \includegraphics[width=\linewidth]{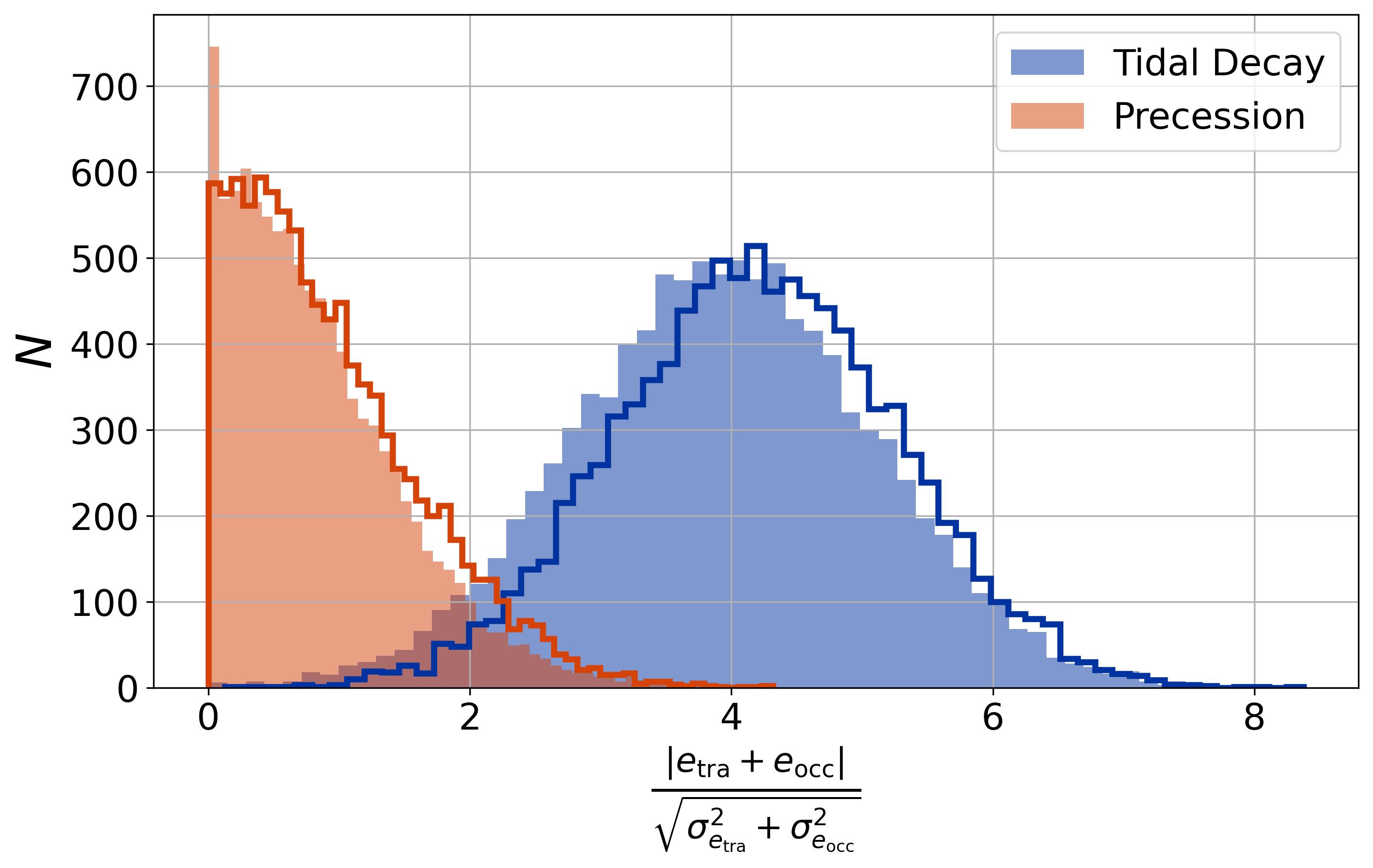}
    \caption{Comparison of the numeric and quasi-analytic eccentricity metrics. For this comparison, we generated 20k synthetic transit and occultation datasets, 10k of them exhibiting tidal decay and 10k of them exhibiting precession. The shaded blue histogram shows the resulting eccentricity metrics assessed numerically, while the blue line shows the metrics assessed quasi-analytically for the tidal decay datasets using Equations \ref{eqn:anal_eccentricity} and \ref{eqn:sigma_anal_eccentricity}. The shaded orange histogram and orange line shows the same statistics but for the precession datasets. The quasi-analytic approach yields metrics slightly larger than the numeric approach, but the mean values agree well within uncertainties.}
    \label{fig:Comparing_Numeric_and_Analytic_Eccentricity_Metrics}
\end{figure}

We checked that there was a reasonable match between the numeric and quasi-analytic approaches by generating 20k synthetic transit- and occultation-timing datasets with observational sampling and Gaussian timing uncertainties equivalent to the dataset shown in Figure \ref{fig:Adding_Occultation_Timing_Data} and with random scatter around a quadratic and a precession ephemeris. For the precession parameters, we used the same values as were used for Figure \ref{fig:Testing_Quad_Fit_to_Precession_Ephemeris}. For the tidal decay parameters, we used the WASP-12 b values. The results of this comparison are shown in Figure \ref{fig:Comparing_Numeric_and_Analytic_Eccentricity_Metrics}. The mean values for the eccentricity metrics from both approaches agree to within uncertainties, as estimated by the standard deviation, though as expected the quasi-analytic approach yielded slightly higher averages. We conducted similar tests using significantly different tidal decay and precession parameters and found good agreement throughout. Of course, for small tidal decay and precessional rates, distinguishing the two signals becomes difficult, and, indeed, the eccentricity metrics for the synthetic tidal decay and precession datasets start to become statistically indistinguishable. 

With these metrics in hand, we next turn to exoplanetary systems that have shown some putative signs of either tidal decay or precession and attempt both to evaluate the quality of the present detections and also explore prospects for future observations.

\section{Application to Additional Systems} \label{sec:Application}

This work was motivated in part by a recent proliferation of claims for non-Keplerian orbital motions that have appeared in the literature. \autoref{table:claims} shows a collection of transiting planets for which some departure from a linear ephemeris has been suggested or detected. Most of these suggestions have arisen in the last five years since long observational baselines are required. Of the eighteen systems for which such a departure has been proposed, only four have had any analysis of occultation-timing data. The others are evenly split between (1) systems for which occultations have been observed but no occultation-timing information has been reported and/or no analysis has been made of the times that have been published; and (2) systems for which no occultations are known to have been observed. In this section, we select two particularly elucidating examples, HAT-P-37 b and WASP-19 b, to demonstrate how the metrics outlined can be applied to decide the robustness of detections of precession or tidal decay based on past observational data. We leave a comprehensive assessment of many more systems to future work.

\begin{table}
   \small
    \centering
    \caption{Planets for which decay has been claimed in the recent literature}
    \begin{tabular}{lcc rc | cc |cc | l}
\hline 
        Planet &        P  & dP/dT  & $\Delta$BIC$^a$ & Ref$^b$ & \multicolumn{2}{c|}{Transits$^c$} & \multicolumn{2}{c|}{Occultations$^c$}  & Comments\\
         &        (d) & (ms yr$^{-1}$) &   & &yr & N& yr & N &  \\
\hline
\multicolumn{10}{c}{\emph{Strong case for decreasing orbital period}} \\ 
        WASP-12 b       & 1.09  &  $-29\pm3~$ & 947$^d$ & [1] & 15  & 102 &  3 & 11  & Likely orbital decay\\
        Kepler-1658 b    & 3.85  & $-131\pm22~$ &  49$^d$ & [2] &  13 &  22 & \multicolumn{2}{c|}{Not used$^e$}  & Plausible decay; single source \\
         WASP-4 b & 1.34 &  $-6.2\pm1.2~$ & 95 & [3] & 13  & 172  & 5 & 4& Cause debated  \\ 
\hline

\multicolumn{10}{c}{\emph{Claim made for possible decreasing period}} \\ 

        HAT-P-32 b & 2.15 & $-30.2\pm3.3~~$ & 7.4$^g$ & [4] & 14 & 87  & \multicolumn{2}{c|}{NA} & Similar precession $\Delta$BIC: 9.4  \\
         HAT-P-53 b & 1.96  &  $-82\pm16~$& 30  & [5] & 4 & 17  &\multicolumn{2}{c|}{NA}& \\
         KELT-9 b & 1.48 & $-24.4\pm 10.7~$ & 8$^h$ &  [6] &  7 & 68  &  6 & 9 & Similar precession $\Delta$BIC: 13 \\
         TrES-1 b & 3.03 & $-16.4 \pm 4.9~$  & 69 & [7] & 19 & 129 & \multicolumn{2}{c|}{Not used$^i$} & $Q_\star^\prime=10^2$  \\ 
         TrES-2 b & 2.47 & $-20.7\pm2.1~~$ & 8 & [4] & 15 & 149 & \multicolumn{2}{c|}{Not used$^j$} & \\
         WASP-43 b & 0.81 & $-5.2 \pm 1.7~$ &  1.5  & [8] & 7 & 56  &  \multicolumn{2}{c|}{Not used} &  Marginal preference for decay \\
        &  & $-1.0 \pm 1.1~$ &   -0.4 & [7]  & 12 & 224  & \multicolumn{2}{c|}{Not used} & No decay  \\
         & & $-1.99\pm0.50~$ & 331 & [9] & 12 & 282 & 12 & 18  & Decay + precession$^k$\\
         XO-2N b  & 2.62 & $-12.95\pm1.85~~$ & 43 & [10]  & 16 & 42 & \multicolumn{2}{c|}{Not used$^l$} &  $Q_\star^\prime=1.5.\times10^3$ \\
         XO-3 b  & 3.19 & $-9\pm5~$ & 383 & [11] [12]  & 12 & 21 & \multicolumn{2}{c|}{Not used$^m$} & $Q_\star^\prime=1.5.\times10^5$ \\
        XO-4 b & 4.2 & $-63\pm17~$ & 13 & [13]  & 15 & 37 & \multicolumn{2}{c|}{NA} & \\

\hline
\multicolumn{10}{c}{\emph{Claim made for possible precessional signal}} \\ 
        HAT-P-37 b& 2.8  & $-80\pm30~$ & 66   & [14] & 10 & 29 & \multicolumn{2}{c|}{NA} &  Stronger precession $\Delta$BIC: 119  \\
\hline

\multicolumn{10}{c}{\emph{Initial claim of decreasing period rebutted by later data}} \\ 
        HAT-P-19 b& 4.01  & $-58\pm7~~$ & 27 & [4] & 12 & 72 & \multicolumn{2}{c|}{NA} &    \\
 &   & NA & -3 & [15] & 12 & 27 & \multicolumn{2}{c|}{NA} & No decay  \\
        HAT-P-51 b & 4.22 & $-135\pm52~~$ & 7 & [5] & 2 & 19 &\multicolumn{2}{c|}{NA} & \\
         &  & NA & -3 & [15] & 2 & 10 &\multicolumn{2}{c|}{NA} & No decay\\
         KELT-16 b & 0.97 & $-27.82\pm 4.76~~$ & 30 &  [13] &  6 & 111  & \multicolumn{2}{c|}{NA} &  \\
        & & $-21\pm 13~$ & 2 &  [7] &  8 & 93  & \multicolumn{2}{c|}{NA} &  Insignificant \\
         TrES-5 b & 1.48 & $-20.4\pm4.7~~$& 14.9$^f$  & [15] & 11 & NA & \multicolumn{2}{c|}{NA} & Similar precession $\Delta {\rm BIC}$:  13.3  \\
         -- & -- & $~-5.3 \pm 2.2~~$& $11\pm20$  & [16] & 18 & 280 & \multicolumn{2}{c|}{NA} & Signal now insignificant  \\
        WASP-19 b & 0.79 & $-6.5 \pm 1.3~$&  209$^n$ &[17][7] & 10 & NA & \multicolumn{2}{c|}{Not used$^o$} & Using literature data to 2022\\
        &  &$0.6 \pm 0.8$ & -4.4 & [7] & 15 & 244 &\multicolumn{2}{c|}{Not used$^o$}  & No decay$^p$ \\
        WASP-32 b & 2.72  &  $-32\pm13~$& 3 & [18] & 11 & 17  &  \multicolumn{2}{c|}{NA} & Marginal  \\
        &   & $-49 \pm 27~$ & 11 &[7] & 14 & 22 & \multicolumn{2}{c|}{NA}  & Insignificant    \\

\hline 

\multicolumn{10}{l}{\small a. $\Delta$BIC $>0$ indicates a quadratic model (e.g., tidal decay) is preferred except as noted.} \\
\multicolumn{10}{l}{\small b. Sources: [1] \citet{2017AJ....154....4P}; [2] \citet{Vissapragada_2022}; [3] \citet{2019AJ....157..217B}; 
 [4] \citet{2022AJ....164..220H}; } \\
\multicolumn{10}{l}{\small ~~~ [5] \citet{2024NewA..10602130Y}; [6] \citet{2023AA...669A.124H};
[7] \citet{2024PSJ.....5..163A}; [8] \citet{2018ChAA..42..101S}; [9] \citet{2025AA...694A.233B};  } \\
\multicolumn{10}{l}{\small ~~~  [10] \citet{2024MNRAS.530.2475Y}; [11] \citet{2022PASP..134b4401Y};  [12] \citet{2023ApJS..264...37S}; [13] \citet{2024ApJS..270...14W};  }\\
\multicolumn{10}{l}{\small ~~~  [14] \citet{2022AJ....163...77A}; [15] \citet{2024MNRAS.528.1930K}; [15] \citet{2021AA...656A..88M}; }\\
\multicolumn{10}{l}{\small ~~~  
[16] Rothmeier et al. (accepted to PSJ);[17] \citet{2020AJ....159..150P}; [18] \citet{2023MNRAS.520.1642S}; } \\
\multicolumn{10}{l}{\small c. Observation stats: yr = years spanned by data used; N = number of unique observations. } \\ 
\multicolumn{10}{l}{\small d. $\Delta$BIC values for WASP-12 b and Kepler-1658b are from \citet{2024PSJ.....5..163A}.} \\ 
\multicolumn{10}{l}{\small e. Occultations are seen in Kepler data but no known timing analysis.} \\ 
\multicolumn{10}{l}{\small i. Occultations exist \citep{2005ApJ...626..523C, 2014ApJ...797...42C} but no known timing analysis.} \\ 
\multicolumn{10}{l}{\small j. Occultations exist, including Kepler \citep{2013ApJ...772...51E}, but no reported timing.} \\ 
\multicolumn{10}{l}{\small k. Fit RV + timing data, prefers combination of tidal decay and apsidal precession; see \citet{2025AA...694A.233B}.} \\ 
\multicolumn{10}{l}{\small l. Four occultation times reported in \citet{2009ApJ...701..514M} but no known timing analysis.} \\ 
\multicolumn{10}{l}{\small m. Twelve occultation times reported in \citet{2014ApJ...794..134W} but no known timing analysis.} \\ 
\multicolumn{10}{l}{\small n. \citet{2020AJ....159..150P} lists no $\Delta$BIC; \citet{2024PSJ.....5..163A} found $\Delta$BIC=209 before removing problem points.} \\ 
\multicolumn{10}{l}{\small o. Occultations exist, including TESS \citep{2020AJ....159..104W}, but no known timing analysis.} \\ 
\multicolumn{10}{l}{\small p. \citet{2024PSJ.....5..163A}  found no evidence for tidal decay after removing problem points from three prior papers.} \\ 
    \end{tabular}
    \label{table:claims}
\end{table}

\subsection{HAT-P-37 b}\label{sec:HAT-P-37b}

\begin{figure}
    \centering
    \includegraphics[width=\linewidth]{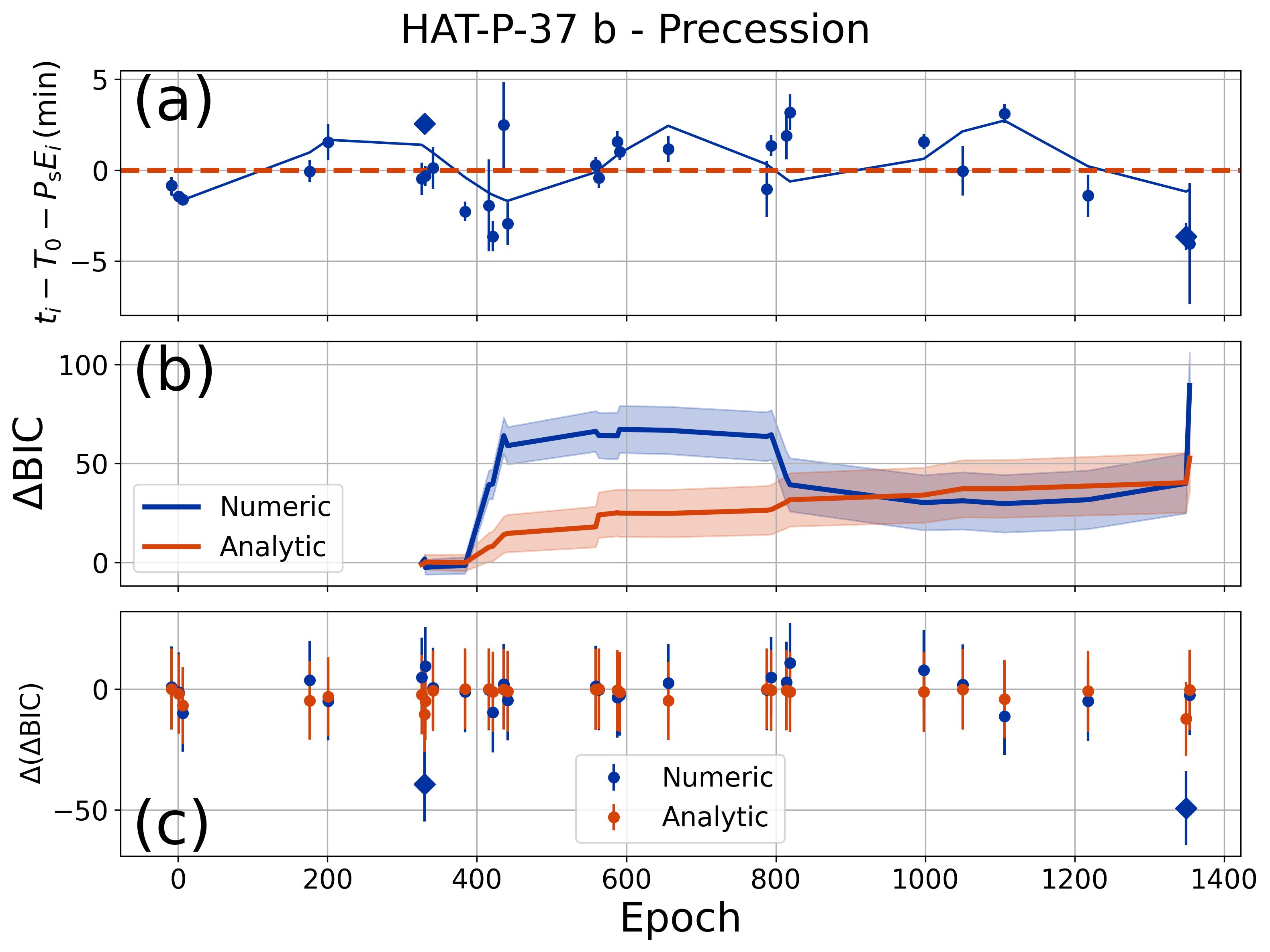}
    \caption{(a) Transit-timing data for HAT-P-37 b as reported in \citet{2022AJ....163...77A}, along with a best-fit precessional model. (b) The running $\Delta {\rm BIC}$ for comparing a linear and a precession model fit to the timing data. The blue curve shows the direct numerical estimate (the blue shaded region, its expected statistical variation), while the orange shows the quasi-analytic estimate using Equation \ref{eqn:BIC_for_lin_fit_to_prec_ephem} (the orange shaded region, its expected variation). Because the precession model requires a minimum of five data points (to fit its five parameters), this calculation does not begin until there are at least five points. (c) The difference in the final $\Delta {\rm BIC}$ value ($\Delta \left( \Delta {\rm BIC}\right)$) that results from removing each timing datum one at a time. The blue points show that calculation using a direct numerical estimate of $\Delta {\rm BIC}$, while the orange points show the expectation using the quasi-analytic estimate (Equation \ref{eqn:BIC_for_lin_fit_to_prec_ephem}). For example, by removing the datum shown by the blue diamond that lies well below the other points near $E = 350$, the final $\Delta {\rm BIC}$ drops by nearly 50\%. The error bars associated with each point correspond to the expected range of statistical variation $\sigma$, and blue points that disagree by more than $1\sigma$ are highlighted in panels (a) and (c) with blue diamonds.}
    \label{fig:OmniPlot_HAT-P-37b_Precession}
\end{figure}

\begin{figure}
    \centering
    \includegraphics[width=\linewidth]{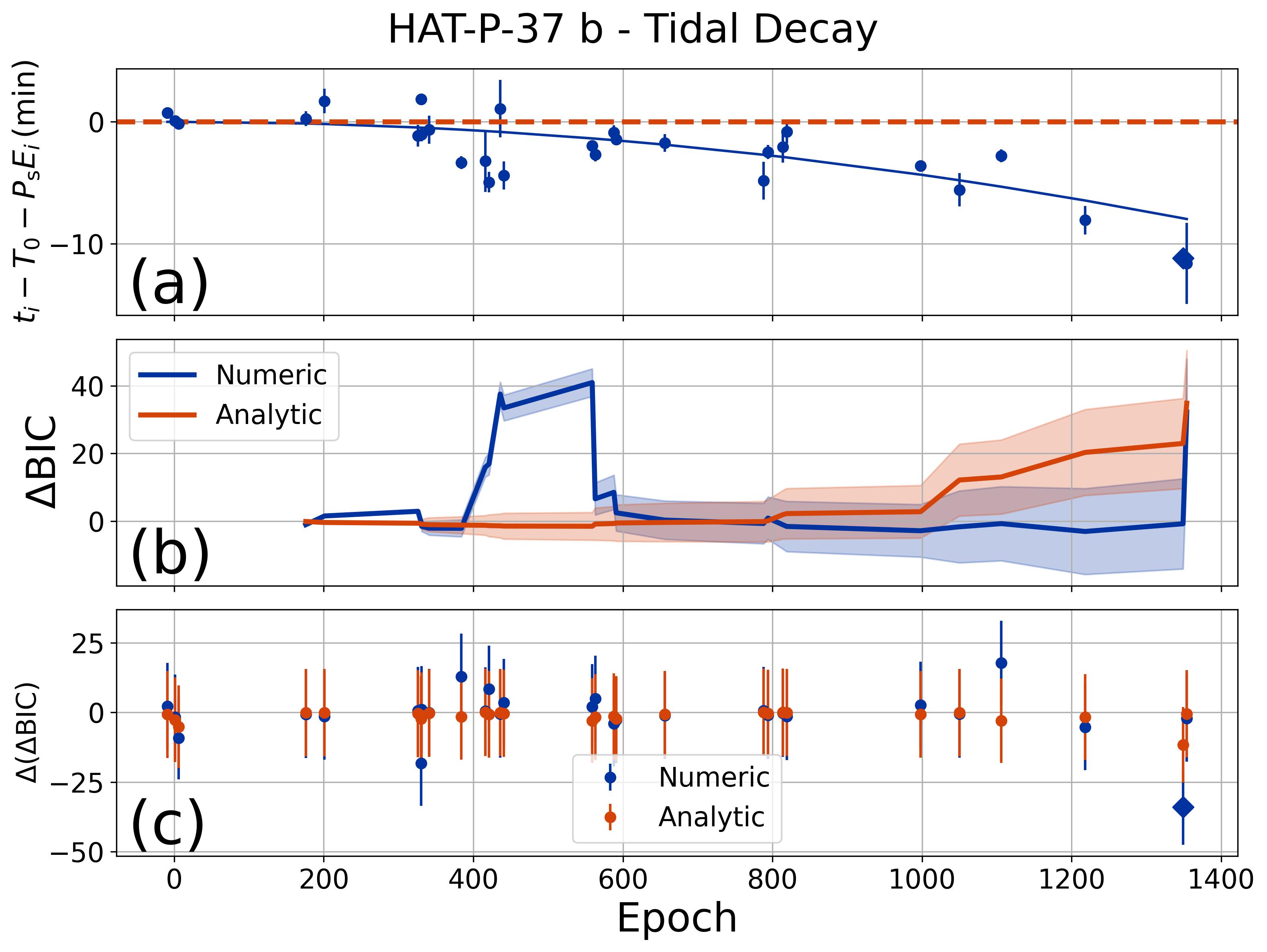}
    \caption{(a) Transit-timing data for HAT-P-37 b as reported in \citet{2022AJ....163...77A}, along with a best-fit tidal decay model. (b) The running $\Delta {\rm BIC}$ for comparing a linear and a tidal decay model fit to the timing data. The blue curve shows the direct numerical estimate (the blue shaded region, its expected statistical variation), while the orange shows the quasi-analytic estimate using Equation \ref{eqn:BIC_for_lin_fit_to_prec_ephem} (the orange shaded region, its expected variation). Because the tidal model requires a minimum of three data points (to fit its three parameters), this calculation does not begin until there are at least three points. (c) The difference in the final $\Delta {\rm BIC}$ value ($\Delta \left( \Delta {\rm BIC}\right)$) that results from removing each timing datum one at a time, normalized to the final $\Delta {\rm BIC}$. The blue points show that calculation using a direct numerical estimate of $\Delta {\rm BIC}$, while the orange points show the expectation using the quasi-analytic estimate (Equation 35 from \citealt{2023AJ....166..142J}). The error bars associated with each point correspond to the expected range of statistical variation $\sigma$, and blue points that disagree by more than one $\sigma$ are highlighted in panels (a) and (c) with blue diamonds.}
    \label{fig:OmniPlot_HAT-P-37b_TidalDecay}
\end{figure}

Discovered by \citet{2012AJ....144...19B}, HAT-P-37 b is a hot Jupiter with a period of about 2.8 days orbiting a 0.9-solar mass G-type star with an effective temperature of 5,500 K. Over a decade of transit observations have yielded a considerable dataset that may show evidence for departure from a linear ephemeris. Analyzing twenty-nine transit times spanning from 2011 to 2021, \citet{2022AJ....163...77A} suggested the planet exhibits precession, perhaps driven by a small (less than Saturn's mass) planet near a 1:2 mean motion resonance with HAT-P-37 b. Panel (a) in both Figures \ref{fig:OmniPlot_HAT-P-37b_Precession} and \ref{fig:OmniPlot_HAT-P-37b_TidalDecay} show the O-C transit-timing data for precession and tidal decay fits, respectively. 

To determine best-fit model parameters, we again apply the Trust Region Reflective algorithm \citep{1982SJNA...19..409S, 2020SciPy-NMeth}. For uncertainties on all fit parameters, we take the square root of the diagonal of the covariance matrix \citep[cf.][]{Press2007}. Fitting the precession model, we retrieve the following best-fit parameters: $T_0 = \left( 2455642.1444 \pm 0.0003 \right)\,{\rm BJD_{\rm TDB} }$, $P_{\rm s} = \left( 2.79744 \pm 0.00005 \right)\,{\rm days}$, $e = 0.001\pm0.0004$, $\omega_0 = \left( -0.7 \pm 0.3 \right)\,{\rm rad}$, and $d\omega/dE = \left( 0.0149 \pm 0.0005 \right)\,{\rm rad\ orbit^{-1}}$. These parameters agree to within uncertainties with the corresponding values from \citet{2022AJ....163...77A}. Fitting the tidal decay model, we retrieve the following best-fit parameters: $T_0 = \left( 2455642.1433 \pm 0.0003 \right)\,{\rm BJD_{\rm TDB} }$, $P_{\rm s} = \left( 2.79744466 \pm 10^{-6} \right)\,{\rm days}$, and $dP/dE = \left( -6 \pm 2 \right)\times10^{-9}\,{\rm days\ orbit^{-1}}$. Again, these parameters agree to within uncertainties with the corresponding values from \citet{2022AJ....163...77A}. 

To determine which model -- linear, precession, or tidal decay -- provides the best-fit to the data, we employ the numerical and the quasi-analytic approaches to explore the evolution of $\Delta {\rm BIC}$ comparing a linear model to both a precession (Figure \ref{fig:OmniPlot_HAT-P-37b_Precession}(b)) and a tidal decay model (Figure \ref{fig:OmniPlot_HAT-P-37b_TidalDecay}(b)). It is important to keep in mind that, for the numerical $\Delta {\rm BIC}$ calculation, as we march forward in time adding datum after datum, we continually update the best-fit parameters for each model, as if we were collecting the observations in real-time. By contrast for the quasi-analytic model, we hold the model parameters at the final best-fit values yielded by the complete dataset under the assumption that the final values are the most accurate reflection of the ``true'' values. This approach allows us gauge which points in the growing dataset have the most impact on the final values.

Looking at panel (b) in both Figures \ref{fig:OmniPlot_HAT-P-37b_Precession} and \ref{fig:OmniPlot_HAT-P-37b_TidalDecay}, we can see the numerical approaches for both models show considerable non-monotonicity in time as the data pile up. The $\Delta {\rm BIC}$ values for both comparisons grow suddenly near $E = 400$ before dropping back to near zero -- for the precession model, near $E = 800$ and for the tidal decay model, near $E = 600$. We see a similar sudden increase in both figures very near the end of the dataset, when $\Delta {\rm BIC}$ in both figures suddenly leaps to large values. Looking at the quasi-analytic calculation in each figure, we see that we would not have expected such large departures. Instead, both figures show $\Delta {\rm BIC}$ would have been expected to have shown slow and steady growth throughout the observational baseline. Thus, we might conclude the transit-timing data harbor some statistical outliers that are artificially driving $\Delta {\rm BIC}$ growth. 

\citet{2024PSJ.....5..163A} proposed a simple approach for determining which data points drive significant excursions in $\Delta {\rm BIC}$. Using transit-timing datasets, that study dropped each datum, one-at-a-time, calculated the resulting $\Delta {\rm BIC}$ for a given model comparison, and then subtracted the final $\Delta {\rm BIC}$ value that arose from using all the data. Here, we call this metric $\Delta \left( \Delta {\rm BIC} \right)$ since it represents a change in $\Delta {\rm BIC}$. Large variations in $\Delta \left( \Delta {\rm BIC} \right)$ indicate that the corresponding datum has outsize influence on the final $\Delta {\rm BIC}$ value. \citet{2024PSJ.....5..163A} flagged points that changed the final $\Delta {\rm BIC}$ by more than 25\% as potentially anomalous under the assumption that no one datum should have a very large impact. 

Panel (c) in Figures \ref{fig:OmniPlot_HAT-P-37b_Precession} and \ref{fig:OmniPlot_HAT-P-37b_TidalDecay} illustrate this calculation. The orange points (labeled ``Analytic'') show how each datum would be expected to impact the final $\Delta {\rm BIC}$ value, with error bars taken as the $\sigma_{\Delta {\rm BIC}}$ values outlined in Section \ref{sec:Analysis}. These points show clearly that, though individual points impact the final $\Delta {\rm BIC}$, none contributes more than about 10-25\%. Compare that result to the $\Delta \left( \Delta {\rm BIC}\right) $ determined via the ``Numeric'' approach. Here, we can clearly see two points in the precession analysis and one point in the tidal decay analysis have outsize influence (shown as blue diamonds in each panel). These points disagree by more than 1$\sigma$ from their corresponding quasi-analytic points. Indeed, if we remove these outlier points from consideration, the final $\Delta {\rm BIC}$ favoring the precession model over a linear model drops from $\Delta {\rm BIC} = 92\pm17$ to $13\pm14$, while the $\Delta {\rm BIC}$ favoring the tidal decay model over a linear model drops from $\Delta {\rm BIC} = 34\pm16$ to $0\pm14$. The $\Delta {\rm BIC}$ comparing for the precession to tidal decay model changes from $-57\pm6$, meaning the precession model is favored, to $64\pm6$, meaning the tidal decay model is favored. Eschewing the outliers also reduces $dP/dE$ to $\left(-2 \pm 3 \right)\times10^{-9}\,{\rm days\ orbit^{-1}}$.

These comparisons highlight how dependent $\Delta\,{\rm BIC}$ may be on a handful of points. Looking at these outliers, the key thing that distinguishes them is unusually small error bars, which may suggest these errors are mis-estimated. Though this result cannot be said to disprove the conclusion that HAT-P-37 b exhibits precession, it certainly raises questions about the robustness of that conclusion.

\subsection{WASP-19 b}\label{sec:WASP-19b}
\begin{figure}
    \centering
    \includegraphics[width=\linewidth]{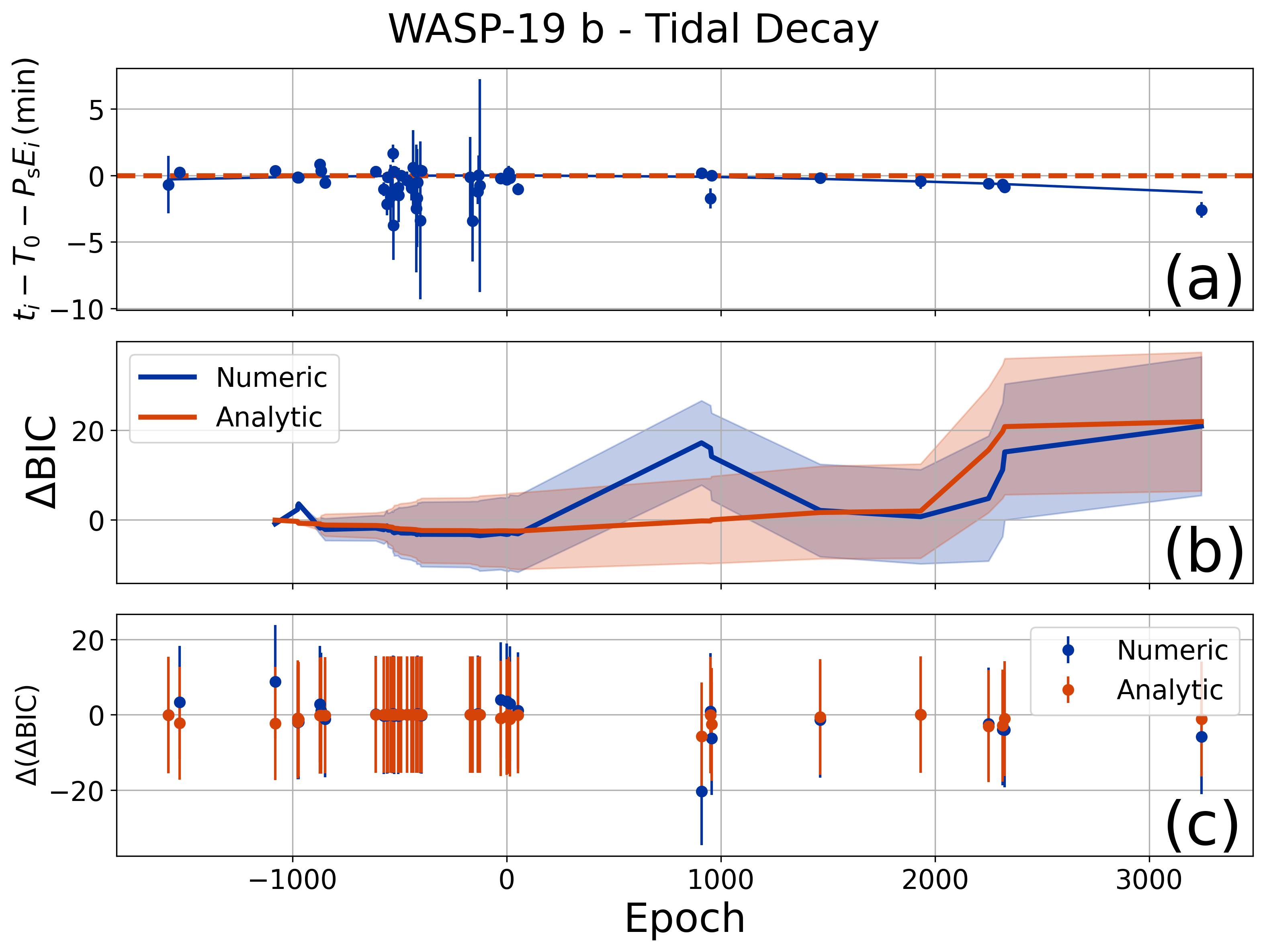}
    \caption{(a) Transit-timing data for WASP-19 b as reported in \citet{2020AJ....159..150P}, along with a best-fit tidal decay model. (b) The running $\Delta {\rm BIC}$ for comparing a linear and a tidal decay model fit to the timing data. The blue curve shows the direct numerical estimate (the blue shaded region, its expected statistical variation), while the orange shows the quasi-analytic estimate using Equation \ref{eqn:BIC_for_lin_fit_to_prec_ephem} (the orange shaded region, its expected variation). Because the tidal model requires a minimum of three data points (to fit its three parameters), this calculation does not begin until there are at least three points. (c) The difference in the final $\Delta {\rm BIC}$ value ($\Delta \left( \Delta {\rm BIC}\right)$) that results from removing each timing datum one at a time, normalized to the final $\Delta {\rm BIC}$. The blue points show that calculation using a direct numerical estimate of $\Delta {\rm BIC}$, while the orange points show the expectation using the quasi-analytic estimate (Equation 35 from \citealt{2023AJ....166..142J}). The error bars associated with each point correspond to the expected range of statistical variation $\sigma$.}
    \label{fig:OmniPlot_WASP-19b_TidalDecay}
\end{figure}

Discovered by \citet{2010ApJ...708..224H}, WASP-19 b seems to have the third shortest orbital period of all confirmed hot Jupiters, 0.78 days. (The Exoplanet Archive lists TOI-2109 as having the shortest period, 0.67 days.) The planet has a mass of about 1.2 Jupiter masses and orbits a main-sequence star similar to the Sun in temperature, mass, and size. Based on fifty-five transit-timing data points, \citet{2020AJ....159..150P} report a possible tidal decay for the planet at a rate of $\left( -2.06 \pm 0.42 \right) \times10^{-10}\,{\rm days\ orbit^{-1}}$, corresponding to a modified tidal quality factor for its host star of $5\times10^5$ -- more dissipative than expected for a solar-type star \citep[e.g.,][]{2024ApJ...960...50W} but similar to the dissipation parameter inferred for WASP-12. There is no evidence or reporting of precession in this system, so for this analysis, we focus on the possibility of tidal decay.

Figure \ref{fig:OmniPlot_WASP-19b_TidalDecay}(a) shows the O-C data for WASP-19 b as reported in \citet{2020AJ....159..150P}. To determine the best-fit tidal decay model, we apply the same approach as we did for HAT-P-37 b and arrive at the best-fit parameters $T_0 = \left( 2456021.70396 \pm 4\times10^{-5} \right)\, {\rm days}$, $P_{\rm s} = \left( 0.78883942 \pm 5\times10^{-8} \right)\,{\rm days}$, and $dP/dE = \left( -1.7 \pm 0.5 \right)\times10^{-10}\,{\rm days\ orbit^{-1}}$. These values agree to within uncertainties with those reported in \citet{2020AJ....159..150P}.

Figure \ref{fig:OmniPlot_WASP-19b_TidalDecay}(b) shows how the evolution for the numerical and quasi-analytic $\Delta{\rm BIC}$ values for comparing the tidal decay to the linear model. The final $\Delta{\rm BIC}$ value for the numerical approach lands on $\left( 21\pm16 \right)$, showing some, though perhaps muted, support for tidal decay over a linear ephemeris. As for HAT-P-37 b, the numerical $\Delta {\rm BIC}$ exhibits significant non-monotonic excursions, rising to nearly 20 at $E = 1000$ before settling back toward zero and then finally climbing to the final value. 

These excursions contrast with the behavior of the quasi-analytic $\Delta {\rm BIC}$, which mostly rises steadily until about $E = 2250$, when it suddenly increases. That increase arises because the timing data at those epochs show the first statistically significant departure from the linear ephemeris (dashed orange line). This increase also manifests in the numerical $\Delta {\rm BIC}$, though it is less pronounced.

Figure \ref{fig:OmniPlot_WASP-19b_TidalDecay}(c) shows no one point appears to be an outlier as measured by the $\Delta \left( \Delta {\rm BIC} \right)$ criterion, although the point near $E = 910$ does appear to have unusually large influence. the impact of each individual data point on the final $\Delta {\rm BIC}$ value, what we have called $\Delta\left(\Delta{\rm BIC}\right)$. Removing that single timing point dramatically reduces the final $\Delta {\rm BIC}$ value to $1\pm9$ and $dP/dE$ to $\left( -8 \pm 5 \right)\times10^{-11}\,{\rm days\ orbit^{-1}}$. 

As with HAT-P-37 b, we see that a small number of points can have outsize impact on the inferred best-fit model. Any time the value for $\Delta\left(\Delta{\rm BIC}\right)$ changes dramatically, it means that a single point may drive detection of non-Keplerian motion, and the odds of a spurious detection may be high. Although this analysis does not in itself disprove tidal decay, it does provide a simple but valuable tool to identify potential problem points. We note that this system was the subject of a re-examination of the literature in \citet{2024PSJ.....5..163A}, which unearthed separate issues in the published transit mid-transit time values provided by several of the sources that were used to compile the time set used by \citet{2020AJ....159..150P}. 

\section{Discussion and Conclusions} \label{sec:Discussion}

\begin{figure}
    \centering
    \includegraphics[width=\linewidth]{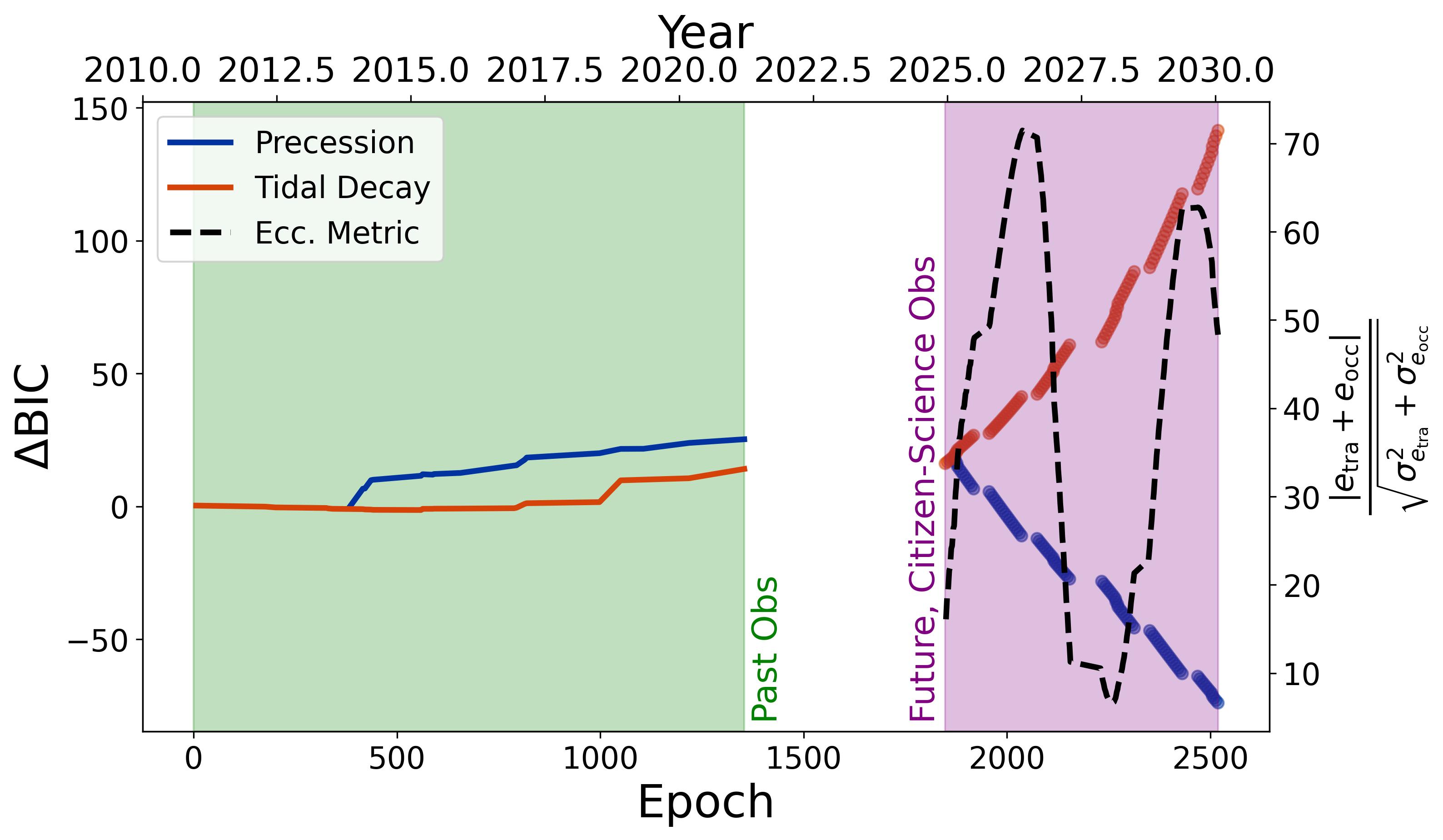}
    \caption{Past and projected future $\Delta {\rm BIC}$ values comparing (1) a precession to a linear model (blue lines and points) and (2) a tidal decay to a linear model (orange lines and points) for the HAT-P-37 b system. At left, the quasi-analytic evolution of $\Delta{\rm BIC}$ is plotted for past observations within the green shaded region; the y-values for these lines should be read off the left axis. Note that we have discarded the outlier points identified in panel (c) in Figures \ref{fig:OmniPlot_HAT-P-37b_Precession} and \ref{fig:OmniPlot_HAT-P-37b_TidalDecay}; not surprisingly, $\Delta{\rm BIC}$ values are lower here, and there is thus less support for a non-Keplerian model. (We have not included the shading showing $\Delta {\rm BIC}$ uncertainties for clarity in this figure.) At right, the blue and orange lines in the purple shaded region show the expected evolution for $\Delta {\rm BIC}$ for a ``Future, Citizen-Science Obs'' campaign as described in Section \ref{sec:Discussion}, also using the quasi-analytic approach. For those calculations, we assumed transit-timing data with poorer precisions $\sigma^{\rm tra} = 10\,{\rm min}$ intended to reflect the kind of precisions accessible for small, citizen-science telescopes. The dashed black line shows evolution of quasi-analytic eccentricity metric (Equation \ref{eqn:anal_eccentricity}), assuming citizen-science transit observations coupled with occultation observations, timing precision for which is assumed to be $\sigma^{\rm occ} = 1.5\,{\rm min}$ as in \citet{2020ApJ...888L...5Y}.}
    \label{fig:HAT-P-37b_Future_Observations}
\end{figure}

As seen in Section \ref{sec:HAT-P-37b}, the conclusion that HAT-P-37 b is experiencing precession hangs on just a handful of data points. Here we will discuss what future observations are needed to robustly detect precession or tidal decay, and describe how to use the metrics presented in this work to optimally plan such a campaign. Figure \ref{fig:HAT-P-37b_Future_Observations} shows both the past and projected future evolution of $\Delta {\rm BIC}$ for comparing both a precession and a tidal decay model to a linear model. For that figure, the green shaded region shows the evolution of the running $\Delta {\rm BIC}$ based on the past transit-timing observations (as shown in Figures \ref{fig:OmniPlot_HAT-P-37b_Precession} and \ref{fig:OmniPlot_HAT-P-37b_TidalDecay}) but after removing the points with unusually large $\Delta \left( \Delta {\rm BIC}\right)$ values (shown as blue diamonds). After removing the outliers identified, the $\Delta {\rm BIC}$ calculated for past data is lower. 

The most obvious solution for resolving whether the system is, indeed, experiencing precession is to collect additional timing data, but how much data, of what quality, and when? Of course, the impact of a planned observational campaign depends very sensitively on the timing precision achievable, among other considerations, but as an illustration, we imagine a future citizen-science-based transit campaign. As a template campaign, we consider results reported in \citet{2023PASP..135a5001P}. That study reported transit data and analysis collected by the Unistellar Exoplanet Campaign, an international collective of citizen scientists in over 61 countries. The campaign used $>$10,000 telescopes of three different, off-the-shelf models, the eVscope 1, eVscope 2, and eQuinox. With apertures of about 10 cm, 281 transits of Jupiter-sized exoplanets orbiting stars with magnitudes $V \le 15.4$ were collected from 2020 to 2022. Though the exact precision on mid-transit times varied, the observations achieved precisions of $\sim 10\,{\rm min}$. Using the metrics outlined in Section \ref{sec:Analysis}, we can explore how such a citizen-science campaign might impact conclusions about HAT-P-37 b's ephemeris.

The purple-shaded region in Figure \ref{fig:HAT-P-37b_Future_Observations} shows the results. For those calculations, we used the astroplan package \citep{astroplan2018} to determine when HAT-P-37 b would be visible from the lead author's host institution. For simplicity, we assumed each visible transit is actually observed (illustrated using the dots). This approach is, of course, overly-optimistic but provides a useful baseline for comparison of other programs. We assumed transit-timing precisions of 10 min and calculated the evolution of $\Delta {\rm BIC}$ for comparing the precession and the tidal decay models both to a linear model. For the blue dots, we assumed HAT-P-37 b exhibits precession with the best-fit parameters reported in Section \ref{sec:HAT-P-37b} and propagated the ephemeris forward in time, calculating the $\Delta {\rm BIC}$ for comparing a precession model to a linear model fit to the resulting times ($\Delta{\rm BIC} > 0$ favors precession). The best-fit precession model for the existing transit-timing data predicts excursions from a linear ephemeris $\Delta t \sim 2\,{\rm min}$. Because these excursions are much smaller than the assumed timing precision of the citizen-science data and because they are oscillatory rather than monotonic, the citizen-science data cannot robustly detect the precession signal. That result is reflected in the fact that the corresponding $\Delta {\rm BIC}$ value going into the future plunges below zero -- $\Delta {\rm BIC}$ arising from the low-precision data would actually favor a linear ephemeris ($\Delta {\rm BIC} < 0$).

For the orange dots, we assumed, instead, that HAT-P-37 b exhibits tidal decay with the best-fit parameters reported in Section \ref{sec:HAT-P-37b} and then calculated the $\Delta {\rm BIC}$ for comparing a tidal decay model to a linear model fit to the resulting mid-transit times ($\Delta {\rm BIC} > 0$ favors tidal decay). That tidal decay model predicts a departure from a linear ephemeris approaching 20 min by 2030 and always in the same direction, meaning the signal builds up over time, rather than oscillating. Consequently, if HAT-P-37 b actually is experiencing tidal decay, we would expect a focused citizen-science transit campaign using 10-cm telescopes could detect the signal. 

Such a campaign coupled to occultation-timing data could distinguish between precession and tidal decay, as shown by the black dashed line in Figure \ref{fig:HAT-P-37b_Future_Observations}. For that calculation, we again assumed HAT-P-37 b exhibits precession and used astroplan to determine when occultations would be visible for HAT-P-37 b from the lead author's home institution and assumed they could all be observed with timing precisions of 1.5 min, comparable to precisions reported in \citet{2020ApJ...888L...5Y}. Of course, these assumptions are overly simplistic and optimistic, but the large values for the eccentricity metric shown in Figure \ref{fig:HAT-P-37b_Future_Observations} suggest even a more limited campaign would likely suffice to distinguish tidal decay from precession, especially if scheduled around the time that the eccentricity metric takes on local maximum values (around $E = 2000$ in early 2027 or later around $E = 2500$ in late 2029). The oscillations in this metric arise from the oscillations expected for a precession model using the best-fit parameters reported above. As the planet's orbit precesses through $\omega = 90^\circ$ ($\cos 90^\circ = 0$ -- see Equation \ref{eqn:transit_precession_ephemeris}), the departures from a linear ephemeris for the transit- and occultation-timing signals become more difficult to detect, reducing the eccentricity metric.

These results illustrate the power of coupling a citizen-science program with professional-grade observational facilities. Collaborations between observational facilities and citizen scientists are already showing the power of people-driven projects. For example, NASA's Exoplanet Watch \citep{2020PASP..132e4401Z} provides participants with data from the Harvard-Smithsonian MicroObservatory \citep{2021JBAA..131..359F} so that anyone, whether or not they own a telescope, can contribute to furthering exoplanet science. A recent study by \citet{2024PASP..136f4401N} of WASP-77 A b specifically showed how the combination of citizen-science and professional data can improve the planet's ephemeris. Further recent work of the Unistellar Exoplanet Campaign leverages their international network and were able to confirm the giant exoplanet TIC 393818343 b \citep{2024AJ....168...26S}.

For the nineteen systems which have been proposed to exhibit a departure from a linear ephemeris as detailed in Table \ref{table:claims}, a combination of archival citizen science data and a new long-term follow-up citizen-science observing campaign would help to ensure that future transits of these planets are being studied. Specifically, in the case of HAT-P-37 b, there are transit observations that have been collected by the MicroObservatory between when the data set used to produce Figures \ref{fig:OmniPlot_HAT-P-37b_Precession} and \ref{fig:OmniPlot_HAT-P-37b_TidalDecay} was published (Jan. 2022) and the submission of this article (April 2025). Combining these observations with planned future campaigns will increase the chance of establishing the cause of the transit-timing variation in HAT-P-37 b. 

Though limited in precision when compared to professional observatories, the power of exoplanet transit observations by citizen scientists lies in their accessibility. A group of amateurs can greatly increase the odds of capturing enough observations to significantly lengthen and fill in observational baselines, meaningfully improving transit ephemerides. Further, because these observations can be done with commercially available telescopes, more transits can be observed without having to propose or compete for telescope time. These improved ephemerides facilitate the detection of non-Keplerian orbital motions such as tidal decay and precession certainly within the observational baselines that were needed for WASP-12 b but potentially even faster. 

One of the key potential concerns for relying on citizen science data -- that it will introduce more low quality data that will be time-consuming to identify and remove from the record  -- is mitigated by the simple metrics that we have developed in this work for identifying points with an outsize effect. Outliers, whether due to observer or publication error or to astronomical oddities (e.g., undetected starspot crossing), already exist in the scientific literature and can affect even the most ostensibly precise data, and no long-term study of timing effects can ignore them. While precession studies may need to restrict themselves to only the more precise transits, searches for tidal decay are well served even by relatively low-precision data so long as they are plentiful and frequent.

\section*{Acknowledgments}
BJ, AK, DC, ERA, AS, and JPM were supported by a grant from the NASA Exoplanets Research Program, award number 80NSSC22K0317. MB was supported by the Idaho Space Grant Consortium. RMH was supported by the NASA Science Activation Program, award number 80NSSC22M0008. The authors acknowledge useful input from Rory Barnes, Eric Agol, and an anonymous referee.

\software{astropy \citep{2013A&A...558A..33A,2018AJ....156..123A}, matplotlib \citep{Hunter:2007}, numpy \citep{harris2020array}, scipy \citep{2020SciPy-NMeth}, astroplan \citep{astroplan2018}}

\appendix

\section{Appendix}\label{sec:Appendix_Fitting_Quadratic_Model_to_Precession_Ephemeris}
Following up on Section \ref{sec:Fitting_a_Quadratic_Model_to_a_Precession_Ephemeris}, we here present the linear regression solution for minimizing $\chi^2$ for fitting a quadratic model to a precession ephemeris. Setting equal to zero the derivatives of $\chi^2$ with respect to each fit parameter, $T_0^\prime$, $P_{\rm s}^\prime$, and $\frac{1}{2} \frac{dP}{dE}$, gives the following system of equations:
\begin{align}
    S_t & = & & S T_0^\prime & + & S_E P_{\rm s}^\prime & + & S_{E^2} \left( \frac{1}{2} \frac{dP}{dE} \right) &\nonumber\\
    S_{Et} & = & & S_E T_0^\prime & + & S_{E^2} P_{\rm s}^\prime & + & S_{E^3} \left( \frac{1}{2} \frac{dP}{dE}\right) &\nonumber\\
    S_{E^2t} & = & & S_{E^2} T_0^\prime & + & S_{E^3} P_{\rm s}^\prime & + & S_{E^4} \left( \frac{1}{2} \frac{dP}{dE} \right), &\label{eqn:quad_ephem_chi_equation_system}
\end{align}
where we have defined $S_\star$ as in Equation \ref{eqn:lin_fit_to_prec_ephem_equation_system}.

As in Equation \ref{eqn:lin_fit_to_prec_ephem_embedded_sums}, the sums involving $t_i$ themselves have sums embedded in them, but the only new one is
\begin{equation}
    S_{E^2t} = S_{E^2} T_0 + S_{E^3} P_{\rm s} - S_{E^2 \cos \omega} \left( \frac{e P_{\rm s}}{\pi} \right). \label{eqn:quad_ephem_embedded_sums}
\end{equation}

We can use the same linear regression approach to solve for $T_0^\prime$, $P_{\rm s}^\prime$, and $\frac{1}{2} \frac{dP}{dE}$:
\begin{align}
    T_0^\prime & = & T_0 + \bigg[\frac{(S_{E^2} S_{E^4} - S_{E^3}^2) S_{\cos \omega} + (S_{E^2} S_{E^3} - S_E S_{E^4}) S_{E\cos \omega} + (S_E S_{E^3} - S_{E^2}^2) S_{E^2\cos \omega}}{\Delta}\bigg] & \left( \frac{e P_{\rm a}}{\pi} \right) &\nonumber\\
    P_{\rm s}^\prime & = & P_{\rm s} + \bigg[ \frac{(S_{E^2} S_{E^3} - S_E S_{E^4}) S_{\cos \omega} + (S S_{E^4} - S_{E^2}^2) S_{E \cos \omega} + (S_E S_{E^2} - S S_{E^3}) S_{E^2 \cos \omega}}{\Delta} \bigg] & \left( \frac{e P_{\rm a}}{\pi} \right) &\nonumber \\
    \left( \frac{1}{2} \frac{dP}{dE} \right) & = & \bigg[ \frac{(S_E S_{E^3} - S_{E^2}^2) S_{\cos \omega} + (S_E S_{E^2} - S S_{E^3}) S_{E \cos \omega} + (S S_{E^2} - S_E^2) S_{E^2 \cos \omega}}{\Delta} \bigg] & \left( \frac{e P_{\rm a}}{\pi} \right) &\label{eqn:quadratic_model_fit_parameter_corrections}
\end{align}
with
\begin{equation}
    \Delta = (S S_{E^2} S_{E^4} + 2 S_E S_{E^2} S_{E^3} - S_{E^2}^3 - S S_{E^3}^2 - S_E^2 S_{E^4}).
\end{equation}
Again, as before, we will collect the terms in square brackets as $\Delta T_0^\prime$, $\Delta P_{\rm s}^\prime$, and $\Delta \left( \frac{dP}{dE} \right)$ corresponding to the corrections for each of the fit parameters, respectively.

\bibliography{sample631}{}
\bibliographystyle{aasjournal}

\end{document}